\documentclass[12pt, a4paper]{article}
\pdfoutput=1

\usepackage{tgtermes}

\usepackage[authoryear]{natbib}
\usepackage{amsmath,empheq,amsfonts,bbm}
\usepackage{epic,eepic,epsfig,longtable}
\usepackage{multirow,verbatim}
\usepackage{subfigure}
\usepackage{array}
\usepackage{wasysym}
\usepackage[ruled,linesnumbered]{algorithm2e}
\usepackage{lineno}

\usepackage[colorlinks,citecolor=blue,linkcolor=blue]{hyperref}
\usepackage{epsfig}
\usepackage{pdflscape}
\usepackage{setspace}
\usepackage{mathrsfs}
\usepackage[figuresleft]{rotating}
\usepackage{amsbsy}
\usepackage{eqnarray}
\usepackage{booktabs}
\usepackage{epstopdf}
\usepackage{color}
\usepackage{cases}
\usepackage{appendix}

\makeatletter
\def\singlespace{\def\baselinestretch{1}\@normalsize}

\makeatletter
\def\singlespace{\def\baselinestretch{1}\@normalsize}

\newcommand{\bfm}[1]{\ensuremath{\mathbf{#1}}}

   \newcommand{\bA}{\bfm A}
   \newcommand{\bB}{\bfm B}  
     
   \newcommand{\bD}{\bfm D}

   \newcommand{\bH}{\bfm H}  
   \newcommand{\bI}{\bfm I}

   \newcommand{\bR}{\bfm R}  \newcommand{\RR}{\mathbb{R}}

   \newcommand{\bU}{\bfm U}  
     
\newcommand{\bw}{\bfm w}   \newcommand{\bW}{\bfm W}  
\newcommand{\bx}{\bfm x}     
   \newcommand{\bY}{\bfm Y}

 \newcommand{\cD}{{\cal  D}}
 \newcommand{\cE}{{\cal  E}}

 \newcommand{\cN}{{\cal  N}}
 \newcommand{\cO}{{\cal  O}}
 \newcommand{\cP}{{\cal  P}}

 \newcommand{\cS}{{\cal  S}}
 \newcommand{\cT}{{\cal  T}}
 \newcommand{\cU}{{\cal  U}}

 \newcommand{\cY}{{\cal  Y}}
 \newcommand{\cZ}{{\cal  Z}}

\newcommand{\bfsym}[1]{\ensuremath{\boldsymbol{#1}}}

              \newcommand{\bGamma}{\bfsym \Gamma}
            \newcommand{\bDelta}{\bfsym {\Delta}}
               
 \newcommand{\bmu}{\bfsym {\mu}}

              \newcommand{\bSigma}{\bfsym \Sigma}
          \newcommand{\bLambda}{\bfsym {\Lambda}}

 \newcommand{\hsigma}{\hat{\sigma}}              
               \newcommand{\hbSigma}{\hat{\bfsym \Sigma}}

\DeclareMathOperator{\argmin}{argmin}

\DeclareMathOperator{\diag}{diag}
\DeclareMathOperator{\E}{E}

\DeclareMathOperator{\Tr}{Tr}

\def \diag   {\mathrm{diag}}
\def \argmin {\mathrm{argmin}}
\def \Cond   {\mathrm{Cond}}
\def \sign   {\mathrm{sign}}

\def \onf    {\mathrm{1,off} }
\def \E      {\mathrm{E}}
\def \med    {\mathrm{med}}

\def \STE    {\tiny\mathrm{STE}}
\def \PDE    {\tiny\mathrm{PDE}}
\def \EC     {\tiny\mathrm{EC}}
\def \t      {\mathrm{\top}}
\def \cO     {\mathcal{O}}

\newcommand{\wbSigma}{\widetilde{\bfsym \Sigma}}

\allowdisplaybreaks
\usepackage{verbatim}
\numberwithin{equation}{section}
\newtheorem{theorem}{Theorem}[section]
\newtheorem{lemma}[theorem]{Lemma}
\newtheorem{condition}{Condition}
\newtheorem{remark}{Remark}
\newtheorem{proposition}{Proposition}
\newtheorem{example}{Example}
\newcounter{CondCounter}
\newenvironment{proof}{{\noindent\it Proof.}\ }{\hfill $\square$\par}

\begin{document}

\title{\textbf{Robust and Well-conditioned Sparse Estimation for High-dimensional Covariance Matrices}\thanks{This work was supported by the National Natural Science Foundation of China (Grant No. 12301346) and the Shandong Province Higher Education Youth Innovation and Technology Support Program (Grant No. 2023KJ199).}
}

\author{Shaoxin Wang \footnote{Corresponding author. School of Statistics and Data Science, Qufu Normal University,  Qufu 273165, China (shxwangmy@163.com or shxwang@qfnu.edu.cn)}, Ziyun Ma.}

\date{\small Last modified on \today}
\maketitle


\begin{abstract}
Estimating covariance matrices with high-dimensional complex data presents significant challenges, particularly concerning positive definiteness, sparsity, and numerical stability. Existing robust sparse estimators often fail to guarantee positive definiteness in finite samples, while subsequent positive-definite correction can degrade sparsity and lack explicit control over the condition number. To address these limitations, we propose a novel robust and well-conditioned sparse covariance matrix estimator. Our key innovation is the direct incorporation of a condition number constraint within a robust adaptive thresholding framework. This constraint simultaneously ensures positive definiteness, enforces a controllable level of numerical stability, and preserves the desired sparse structure without resorting to post-hoc modifications that compromise sparsity. We formulate the estimation as a convex optimization problem and develop an efficient alternating direction algorithm with guaranteed convergence. Theoretically, we establish that the proposed estimator achieves the minimax optimal convergence rate under the Frobenius norm. Comprehensive simulations and real-data applications demonstrate that our method consistently produces positive definite, well-conditioned, and sparse estimates, and achieves comparable or superior numerical stability to eigenvalue-bound methods while requiring less tuning parameters.
\end{abstract}


\section{Introduction}
\label{sec:intro}

In the era of big data, covariance matrix estimation has emerged as a fundamental challenge in modern statistical inference, particularly in high-dimensional settings where the dimensionality $p$ rivals or exceeds the sample size $n$. Traditional methods, notably the sample covariance matrix that has served as a cornerstone in multivariate analysis for decades, exhibit significant limitations in these scenarios. These limitations include rank deficiency when $p > n$, degraded estimation accuracy due to the accumulation of errors across $\mathcal{O}(p^2)$ parameters \citep{MarcP67, JohnL09}, and limited structural interpretability in complex data environments. The ubiquity of high-dimensional data across diverse fields such as genomics, financial econometrics, and machine learning has intensified the demand for innovative estimation frameworks that can effectively address the curse of dimensionality while maintaining essential statistical properties. Recent advances, such as thresholdings \citep{BickL08, EKaro08, RothLZ09, CaiL11, CaiY12}, shrinkages \citep{LedoW04, SchaS05, HuanF19, WonLKR13, LedoW22}, and structure assumptions \citep{WuP03, Pour13, FanFL08, FanLM11, FanLM13},  have made tremendous contributions to address these challenges.

Thresholding techniques, in particular, have attracted considerable attention due to their computational efficiency and strong asymptotic properties. Under the assumption that the columns or rows of the true covariance matrix belong to an $l_q$-ball and the data follows a sub-Gaussian distribution, consistent estimates can be obtained through thresholding \citep{BickL08, RothLZ09, CaiL11}. However, these methods face challenges in ensuring positive definiteness in finite samples, potentially yielding indefinite estimates \citep{Roth12, XueMZ12, LiuWZ14}. To address these limitations, researchers have introduced positive definite constraints \citep{BienT11, XueMZ12, LiuWZ14} and log-determinant barrier functions \citep{Roth12} into the covariance matrix estimation framework.
Though these modifications improve the positive definiteness of the estimates by giving a small lower bound on its eigenvalues, they also introduce new challenges in parameter selection and may result in numerical instability \citep{LedoW04,WonLKR13,WangY21}.
A particularly notable application of these methods is in the Markowitz portfolio optimization problem, where ill-conditioned covariance matrix estimates can significantly impact the quality of investment strategies \citep{Mark52,Guig11,WangY21}.

The dispersion of eigenvalues in sample covariance matrices, as documented by \citep{MarcP67, LedoW04}, presents another significant challenge, with largest eigenvalues biased upwards and smallest eigenvalues biased downwards. This phenomenon results in ill-conditioned estimates, often leading to numerical instability in the following model computations. Shrinkage techniques and condition number constraints have emerged as effective solutions, not only ensuring positive definiteness but also enhancing numerical stability \citep{LedoW04, WangP15, WonLKR13}. The condition number constrained maximum likelihood estimation (MLE) approach, in particular, has demonstrated superior performance by truncating extreme eigenvalues from both ends \citep{WonLKR13,AubrDPF12}, finding applications in diverse fields including digital signal processing \citep{AubrDPF12,PanYY17}, portfolio management \citep{DeshD20} and so on.

The sub-Gaussian assumption, while theoretically convenient, significantly limits the applicability of high-dimensional covariance matrix estimation methods to more complex data settings. This limitation has spurred substantial research into robust estimation frameworks that can handle outliers and heavy-tailed distributions \citep{ChenWH11, OlleC15, AvelBFL18, ChenGR18, LohT18, PacrL23}. The pairwise approach, which constructs robust sample covariance matrices by capturing the covariance or correlation between pairs of variables, has emerged as a particularly promising direction \citep{OlleC15, AvelBFL18}.
Under some moment assumptions, \cite{AvelBFL18} applied the adaptive thresholding technique to the robust pilot estimators, robust counterparts of sample covariance matrix, and proposed the robust adaptive thresholding covariance matrix estimator (RATE) to handle leptokurtic data, achieving minimax convergence rates comparable to those under sub-Gaussian assumptions \citep{CaiL11}. However, these robust estimators still face challenges in maintaining positive definiteness in finite samples, often requiring projections onto the cone of positive semidefinite matrices that may compromise sparsity structures.

Based on these achievements, we propose a Robust and Well-conditioned Sparse (RWS) covariance matrix estimation method that incorporates condition number constraints in high-dimensional complex data settings. Our method represents a well-conditioned modification of the robust adaptive thresholding covariance matrix estimator proposed by \cite{AvelBFL18}, while drawing inspiration from the contributions of \citep{XueMZ12, LiuWZ14, CuiLS16}. The proposed approach addresses key limitations in existing methods by maintaining positive definiteness, ensuring numerical stability, and preserving sparsity structures, while requiring fewer tuning parameters than comparable methods \citep{LiuWZ14}.

Before going to the details, we first introduce the following notation used through out the paper.  For real numbers $\alpha$ and $\beta$, $\sign(\alpha)$ gives the sign of $\alpha$, $\alpha_+ = \max\left\{\alpha, 0\right\}$, and $[\alpha, \beta]$ denotes the interval from $\alpha$ to $\beta$ with $\alpha\leq\beta$.  For matrices $\bA = [a_{ij}]$ and $\bB = [b_{ij}]\in \RR^{p\times p}$, $\bA^{\t}$ denotes the transpose of $\bA$, $\gamma_{\max}(\bA)$ and $\gamma_{\min}(\bA)$ are the largest and smallest eigenvalues of $\bA$, and $\langle\bA,\bB\rangle = \Tr(\bA^{\t}\bB)$. We use the following matrix norms: $\|\bA\|_F = \sqrt{\Tr(\bA^{\t}\bA)} = \sqrt{\sum_{i,j}a_{ij}^2}$, $\|\bA\|_{\onf} = \sum_{i\neq j}|a_{ij}|$, and $\|\bA\|_2 = \sqrt{\gamma_{\max}(\bA^{\t}\bA)}$. $\cP_{+}$ and $\cP_{++}$  denote the sets of positive semidefinite and positive definite matrices, respectively. When $\bA$ and $\bB\in\cP_{+}$, we use $\bA\succeq \bB$ to denote the matrix $\bA-\bB$ is positive semidefinite. When $\bA\in\cP_{++}$, the condition number of $\bA$ is defined by $\Cond(A) = \gamma_{\max}(A)/\gamma_{\min}(A)$.

\section{The RWS estimation method}

In this part, we will give the details of our method, which shows that the RWS estimation of covariance matrix achieves well-conditioned and sparse estimation simultaneously rather than performing post modification on the RATE \citep{AvelBFL18}. The methodology is similar to \cite{XueMZ12}, but with more focus on numerical stability and more complex data settings.

\subsection{Related works and motivation}

Let $X$ be a $p$-dimensional random vector with mean $\bmu$ and covariance matrix $\bSigma^* = \E (X - \bmu)(X-\bmu)^{\t}$. The sample covariance matrix is defined as $\widehat{\bSigma}_n = {1}/{n}\sum_{i=1}^{n}(\bx_i- \bar{\bx})(\bx_i - \bar{\bx})^{\t}$,
where $\bx_i$ is the observations of $X$ and $\bar{\bx} = 1/n\sum_{i=1}^n\bx_i$. According to \citep{RothLZ09},  the soft-thresholding estimator (STE) of $\bSigma^*$ is given by
\begin{equation}\label{eq.STE}
  \widehat{\bSigma}_{\STE} =\cS_{\lambda}\left(\widehat{\bSigma}_n\right)= \left[s_{\lambda}(\hat{\sigma}_{ij})\right]
=\left\{\begin{array}{ll}
   \sign(\hat{\sigma}_{ij})\left(|\hsigma_{ij}| - \lambda\right)_+, & \hbox{$i \neq j$,} \\
   \sigma_{ij}, & \hbox{$i = j$,}
   \end{array}
  \right.
\end{equation}
which is equivalent to solving the $L_1$ penalized optimization problem
\begin{equation}\label{eq.opt1}
  \widehat{\bSigma}_{\STE} = \mathop\argmin_{\bSigma}\frac{1}{2} \left\|\bSigma - \widehat{\bSigma}_n\right\|_F^2 + \lambda\left\|\bSigma\right\|_{\onf}.
\end{equation}
Under Gaussian and approximate sparse assumptions, \cite{RothLZ09} established the consistency and sparsistency properties, which implies with an overwhelming probability $\widehat{\bSigma}_{\STE}$ is positive definite and has the right sparse structure. However, \cite{XueMZ12} showed that with finite sample the STE can be indefinite, so they proposed to add the positive definite constraint $\bSigma\succeq \tau\bI$ to \eqref{eq.opt1} and got the following positive definite covariance estimator (PDE)
\begin{equation}
\label{eq.opt2}
  \widehat{\bSigma}_{\PDE} = \mathop\argmin_{\bSigma\succeq \tau\bI}\frac{1}{2} \left\|\bSigma - \widehat{\bSigma}_n\right\|_F^2 + \lambda\left\|\bSigma\right\|_{\onf},
\end{equation}
where $\tau > 0$ serves as the lower bound on the eigenvalues of $\widehat{\bSigma}_{\PDE}$.
With same consideration, \cite{LiuWZ14} proposed correlation-based estimator with nonconvex penalty and eigenvalue constraints, denoted by EC2. Different from \cite{XueMZ12}, \cite{LiuWZ14} gave more focus on the statistical properties especially the minimax optimal convergence rate. It should be noted that a closely related form given in \citep{LiuWZ14} is
\begin{equation}\label{eq.opt3}
  \widehat{\bSigma}_{\EC 2} = \mathop\argmin_{\tau_2\bI\succeq\bSigma\succeq \tau_1\bI} \frac{1}{2}\left\|\bSigma - \widehat{\bSigma}_n\right\|_F^2 + \lambda\left\|\bSigma\right\|_{\onf},
\end{equation}
where $\tau_2>0$ and $\tau_1>0$ are upper and lower bounds on the eigenvalues of $\widehat{\bSigma}_{\EC 2}$. The EC2 method employs two parameters to perform truncation of the eigenvalues of $\widehat{\bSigma}_{\EC 2}$, whereas we will show that our method is more numerical stability oriented and only use one parameter to achieve the truncation process. The statistical properties of the aforementioned works are mainly established under sub-Gaussion assumption. When the data contains contaminations or is from heavy-tailed distributions, the violation of sub-Gaussion assumption may give distorted results (see e.g. \cite{OlleC15,LohT18,AvelBFL18,PacrL23}).

It is widely recognized that the sensitivity of sample covariance matrices to outliers renders existing covariance matrix estimation methods ineffective in producing reliable results.
The pairwise approach can largely reduce the influence of outliers and are often used to construct robust covariance matrix estimators \citep{OlleC15,LohT18,AvelBFL18}. As the robust counterpart of $\widehat{\bSigma}_n$, \cite{AvelBFL18} proposed to construct the following robust pilot estimator
$\wbSigma_n = \left[\tilde{\sigma}_{ij}\right]$ constructed using three different pairwise correlation/covariance estimators:
\begin{enumerate}
  \item \emph{Rank-based} estimator. Suppose the covariance matrix $\bSigma^* = \bD^*\bR^*\bD^*$ with $\bR^* = \left[r_{ij}\right]$ denoting the correlation matrix and $\bD^* = \diag (({\sigma}_{11}^*)^{1/2},\cdots, ({\sigma}_{pp}^*)^{1/2})$, then the rank-based estimator $\wbSigma_n^{R}$ is constructed as follows
\begin{equation*}
  \wbSigma_n = \widetilde{\bD}\widetilde{\bR}\widetilde{\bD}
\end{equation*}
with $\widetilde{\bD} = \diag\left((\tilde{\sigma}_{11})^{1/2},\cdots, (\tilde{\sigma}_{pp})^{1/2}\right)$ and $\widetilde{\bR} = \left[\tilde{r}_{ij}\right]$. Here $\tilde{\sigma}_{uu} = C_u\med_{i\in \{1,\cdots, n\}}\{|x_{iu} - \med_{j\in \{1,\cdots, n\}}(x_{ju})|\}$
 is the median absolute deviation estimator of the ${\sigma}_{ii}$, $\med_{i\in \{1,\cdots, n\}}(\cdot)$ is to take the median with the index set $\{1,\cdots, n\}$, and $C_u$ is the Fisher consistency constant. $\tilde{r}_{ij} = \sin(\pi\tilde{\tau}_{ij}/2)$ with $\tilde{\tau}_{ij}$ being the empirical Kendall's tau correlation between $X_i$ and $X_j$.
  \item \emph{Adaptive Huber} estimator. By the equality $\sigma_{ij}^* = \mu_{ij}^* - \mu_i^*\mu_j^*$ with $\mu_{ij}^* = \E(X_iX_j)$, $\mu_i^* = \E X_i$ and $\mu_j^* = \E X_j$, the robust estimation of $\sigma_{ij}^*$ can be achieved by the robust estimation of the mean values $\mu_{ij}^*$, $\mu_i^*$ and $\mu_j^*$. \cite{AvelBFL18} adopted the following adaptive Huber M-estimator \citep{FanLW17} of mean $\mu$
\begin{equation*}
  \sum_{k = 1}^n\psi_H(z_k - \mu) = 0
\end{equation*}
to give the robust estimates $\tilde{\mu}_{ij}^H$, $\tilde{\mu}_i^H$ and $\tilde{\mu}_j^H$ by substituting $z_i$ with $x_{ik}x_{jk}$, $x_{ik}$ and $x_{jk}$, respectively. The parameter $H$ provides the flexibility to control the trade-off between bias and robustness in high-dimensional data setting.
  \item \emph{Median of means} estimator. In a similar manner to the adaptive Huber M-estimator,  the median of means estimator is given by $\wbSigma_n^{M}=\left[\tilde{\sigma}_{ij}^{M}\right] = \left[\tilde{\mu}_{ij}^{M} - \tilde{\mu}_i^{M}\tilde{\mu}_j^{M}\right]$, where $\tilde{\mu}_{ij}^{M}$, $\tilde{\mu}_i^{M}$ and $\tilde{\mu}_j^{M}$ are the medians of means estimators of $\left\{x_{ik}x_{jk}\right\}_{k=1}^n$, $\left\{x_{ik}\right\}_{k=1}^n$ and $\left\{x_{jk}\right\}_{k=1}^n$ respectively, $M$ is the number of groups into which the dataset is divided.
\end{enumerate}
Building upon these robust pilot estimators, \cite{AvelBFL18} applied adaptive thresholding technique to $ \wbSigma_n$, obtaining
\begin{equation*}
    \wbSigma_n^{\cT} = \left[s_{\lambda_{ij}}(\tilde{\sigma}_{ij})\right],
\end{equation*}
where $\lambda_{ij} = \lambda\left(\tilde{\sigma}_{ii}\tilde{\sigma}_{jj}\log p/n\right)^{1/2}$ is an entry-dependent threshold with $\lambda>0$. Under mild conditions, \cite{AvelBFL18} established that the aforementioned pilot estimators satisfy Condition~\ref{Lemma2}. This condition enables the derivation of minimax convergence rates and was subsequently employed by \cite{WangKW24} in developing non-convex sparse estimators for high-dimensional covariance matrices.
\begin{condition}
\label{Lemma2}
Under mild regularity conditions, the pilot estimator $\wbSigma_n$ satisfies
\begin{equation*}
\label{eq.lema2}
\mathrm{Pr}\left(\max_{i,j}\left|\tilde{\sigma}_{ij} - \sigma^*_{ij}\right|\leq C_0\sqrt{\frac{\log p}{n}}\right)\geq 1- \epsilon_{n,p},
\end{equation*}
where $C_0$ is a positive constant and $\epsilon_{n,p}$ is a deterministic sequences converges to 0 as $n, p\rightarrow \infty$.
\end{condition}

\begin{figure}[htp]
  \centering
  \includegraphics[width=0.85\textwidth,height=0.43\textwidth]{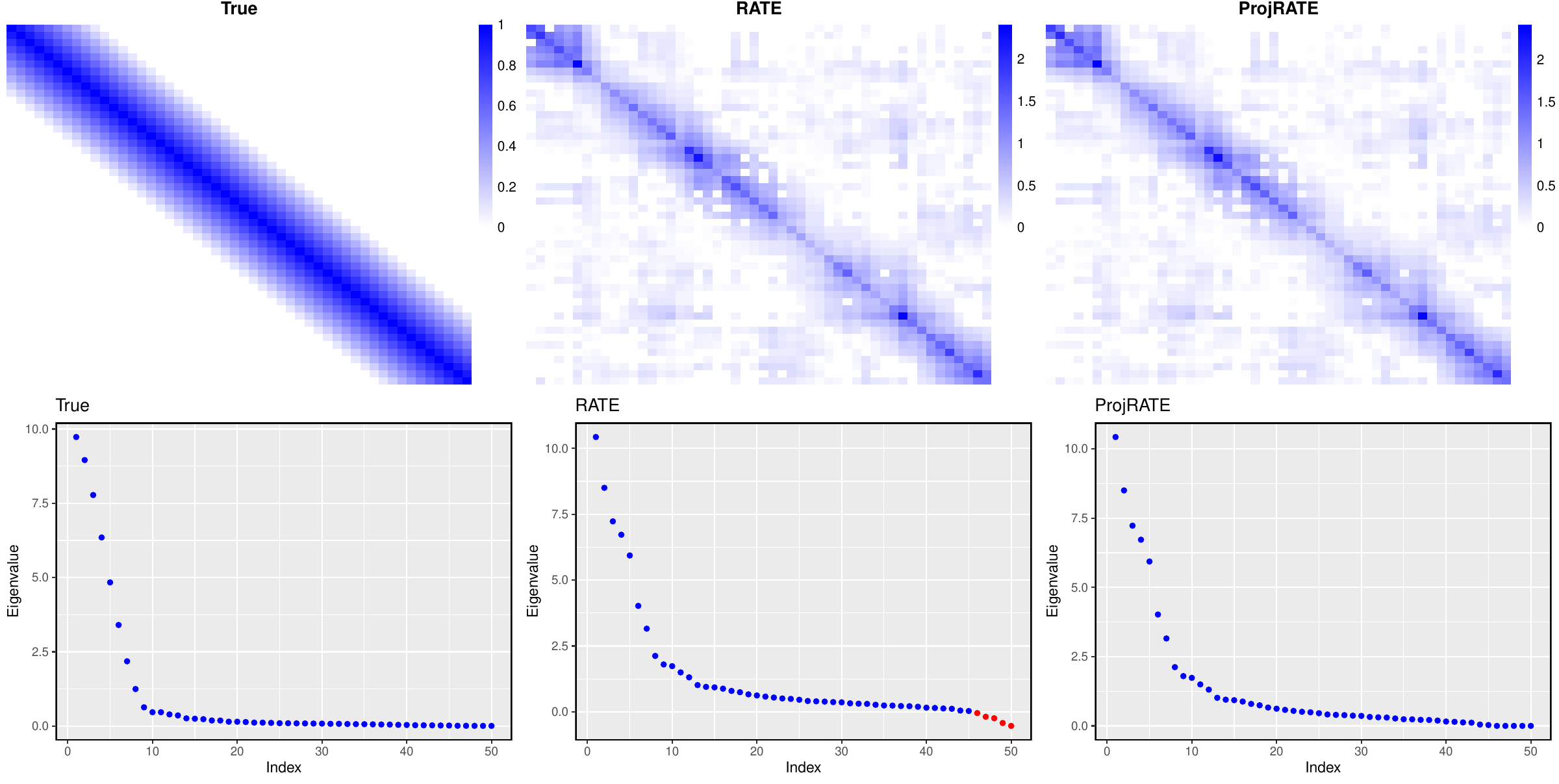}\\
  \caption{Heatmaps and eigenstructure of the true covariance matrix and its RATE and ProjRATE estimates, with negative eigenvalues highlighted in red.}\label{Fig1}
\end{figure}

However, in the finite sample setting, $ \wbSigma_n^{\cT}$ may be indefinite and \cite{AvelBFL18} suggested that the positive semidefinite modification can used. Here we present a toy example in Figure~\ref{Fig1}, which has the same setting as that in Example~\ref{example1} except for $n = 50, p = 50$. For reproducibility, the data is randomly generated with a prescribed seed number \texttt{12345}.
From Figure~\ref{Fig1}, we note that the RATE not only has a diverging eigenstructure but also produces non-negligible negative eigenvalues. Furthermore, we also find that the true covariance matrix $\bSigma^*$ has a sparsity with 1640 zeros and $\wbSigma_n^{\cT}$  has 1196 zeros, whereas the positive semidefinite modification of $\wbSigma_n^{\cT}$ only has 788 zero elements. This implies that the positive semidefinite modification significantly degrades the sparsity structure of $\wbSigma_n^{\cT}$. Most recently, to ensure the positive definiteness of RATE, \cite{WangKW24} proposed a positive definite constrained and nonconvex regularized high-dimensional covariance matrix estimator, and established the theoretical properties of the estimator and its numerical solver, which can also be treated as the generalization of \cite{XueMZ12} to nonconvex penalty and more complex data settings. Different from the aforementioned works and considering the dispersion property of the eigenstructure of RATE, we proposed the following robust high-dimensional covariance matrix estimator with more focus on the numerical stability.

\subsection{The proposed method}

To remove the influence of overestimation of the extreme eigenvalues in covariance matrix estimation, we introduce the condition number constraint which not only ensures the positive definiteness but also gives a good control of the numerical stability. Our proposed RWS covariance matrix estimator is given as follows
\begin{eqnarray}
\label{eq.propmodel}
\hat{\bSigma} &=& \mathop\argmin_{\bSigma\in \cP_{+}} \frac{1}{2}\left\|\bSigma - \wbSigma_n\right\|^2_F + \lambda\left\|\bSigma\right\|_{\onf}, \quad\textrm{s.t.}\;
\mathrm{Cond}\left(\bSigma\right)\leq \kappa_n,
\end{eqnarray}
where $\lambda > 0$ is the tuning parameter controlling the sparsity of $\hat{\bSigma}$ and $\kappa_n \in [1,\infty)$ is the upper bound of the condition number of $\hat{\bSigma}$. The constraint and the penalization term in the proposed RWS estimator can simultaneously guarantee that $\hat{\bSigma}$ is positive definite, preserves controllable numerical stability, and has a sparse structure. Compared with \eqref{eq.opt3} proposed by \cite{LiuWZ14}, our method needs less tuning parameters but achieve the same numerical performance. Moreover, it can be easily proved that the optimization problem \eqref{eq.propmodel} is convex and can be efficiently solved.

\begin{lemma}
\label{Lemma1}
The optimization problem \eqref{eq.propmodel} is convex.
\end{lemma}
\begin{proof}
See Appendix~\ref{AppendA}.
\end{proof}

The convexity of \eqref{eq.propmodel} ensures $\hat{\bSigma}$ is the global minimizer. To study its statistical property, we also need the following condition.
\begin{condition}
\label{Condition2}
The matrix $\bSigma^*$ belong to the class
\begin{equation*}
\label{eq.lema2}
\cU_0 = \left\{\bSigma: \max_{i}\sum_{j=1}^p |\sigma_{ij}^*|\leq s, \bSigma\in \cP_{+} \textrm{ and } \mathrm{Cond}\left(\bSigma\right)\leq \kappa_n\right\}.
\end{equation*}
\end{condition}
With Conditions~\ref{Lemma2} and \ref{Condition2} and similar to \citep{AvelBFL18, WangKW24}, we can also establish the rate of convergence of RWS covariance matrix estimator and present it in the following theorem.
\begin{theorem}
\label{Thm1}
Suppose the pilot estimators $\wbSigma_n$ satisfy Condition~\ref{Lemma2} and  $\lambda = C_0\sqrt{\log p/n}$, then the RWS estimator $\hbSigma$ satisfies
\begin{eqnarray*}
\mathrm{Pr}\left(\left\|\hbSigma - \bSigma^{*}\right\|_F \leq C_2\sqrt{\frac{(s + p)\log p}{n}}\right)\geq 1- \epsilon_{n,p},
\end{eqnarray*}
where $C_2$ is some positive constant.
\end{theorem}
\begin{proof}
See Appendix~\ref{AppendB}.
\end{proof}

Theorem~\ref{Thm1} shows that for any pilot estimator $\wbSigma_n$ satisfying Condition~\ref{Lemma2} the RWS estimator achieve the convergence rate $\cO_p(\sqrt{(s + p)\log p/ n})$ under Frobenius norm. Though the RWS estimator can be treated as a direct generalization of the results given by \cite{XueMZ12} and \cite{LiuWZ14} to more complex heavy-tailed data settings, it achieves the same convergence rate for covariance matrix estimation as that in sub-Gaussian data settings.

\subsection{The alternating direction algorithm}
\label{sec:algorithm}
The alternating direction method (ADM) can be used to establish an iterative framework that strategically decouples the optimization problem \eqref{eq.propmodel} into multiple subproblems, each of which can be efficiently solved \citep{Boyds11}. To apply the ADM, we first introduce an auxiliary variable $\bY$ and rewrite the optimization problem \eqref{eq.propmodel} in the following equivalent form
\begin{eqnarray}
\label{eq.eqvamodel}
\left(\hat{\bSigma}^{*},\hat{\bY}^{*}\right) &=& \mathop\argmin_{\bSigma,\bY\in \cP_{+}} \frac{1}{2}\left\|\bSigma - \wbSigma_n\right\|^2_F + \lambda\left\|\bSigma\right\|_{\onf}, \quad\textrm{s.t.}\;
 \left\{
   \begin{array}{ll}
     \bY = \bSigma, & \hbox{ } \\
     \Cond\left(\bY\right)\leq \kappa_n. & \hbox{ }
   \end{array}
 \right.
\end{eqnarray}
Then the augmented Lagrangian function of \eqref{eq.eqvamodel} is given by
\begin{eqnarray*}
\label{eq.lagrange}
  L_{\mu}\left(\bSigma,\bY,\bLambda\right) &=& \left\{ \frac{1}{2}\left\|\bSigma - \wbSigma_n\right\|^2_F + \lambda\left\|\bSigma\right\|_{\onf} + \left\langle\bSigma-\bY,\bLambda\right\rangle+ \frac{1}{2\mu}\left\|\bSigma - \bY\right\|_F^2\right\},\\
& \textrm{s.t.}&
 \left\{
   \begin{array}{ll}
     \bY = \bSigma, & \hbox{ } \\
     \Cond\left(\bY\right)\leq \kappa_n, & \hbox{ }\\
     \bY \in \cP_{+}, & \hbox{ }
   \end{array}
 \right.\nonumber
\end{eqnarray*}
where $\bLambda$ is the Lagrangian multiplier matrix and $\mu$ is the penalty parameter. Based on the iterative framework of ADM,  suppose we have obtained the $k$-th iteration results $\bSigma^{k}$, $\bY^{k}$ and $\bLambda^{k}$, then the $k+1$-th iteration is performed with the following iterative procedure
\begin{subequations}
\label{eq.lags}
\begin{empheq}[left=\empheqlbrace]{align}
  \bY^{k+1} &= \mathop\argmin_{\bY \in \cP_{+},\Cond\left(\bY\right)\leq \kappa_n} L_{\mu}\left(\bSigma^{k},\bY,\bLambda^{k}\right), \label{eq.lag2}\\
\bSigma^{k+1}& = \mathop\argmin_{\bSigma} L_{\mu}\left(\bSigma,\bY^{k+1},\bLambda^{k}\right), \label{eq.lag1}\\
  \bLambda^{k+1}& = \bLambda^{k} + \frac{1}{\mu}\left(\bSigma^{k+1} - \bY^{k+1}\right).\label{eq.lag3}
\end{empheq}
\end{subequations}
%
This iterative process continues until a predefined convergence criterion is met. We now to show the subproblems \eqref{eq.lag1} and \eqref{eq.lag2} can be very efficiently solved.

To update the auxiliary variable $\bY$, from \eqref{eq.lag2} we have
\begin{equation}
\label{eq.lag22}
  \bY^{k+1} = \mathop\argmin_{\bY \in \cP_{+},\Cond\left(\bY\right)\leq \kappa_n} \left\|\bY - \cY\right\|_F^2
\end{equation}
with $\cY = \bSigma^{k} + \mu\bLambda^{k}$.
Unlike the eigenvalue constraints specified in \eqref{eq.opt2} and \eqref{eq.opt3}, the condition number constraint $\Cond\left(\bY\right)\leq \kappa_n$ prevents \eqref{eq.lag22} from having a closed-form solution. Nevertheless, we can demonstrate that \eqref{eq.lag22} can be solved with a computational burden comparable to that of the aforementioned eigenvalue constraints with the following result.

\begin{proposition}
\label{Prop1}
Suppose $\cY = \bU\widehat{\bGamma}\bU^{\t}$ is the spectral decomposition of $\cY$ and $\widehat{\bGamma}$ is a diagonal matrix with entries $\hat{\gamma}_1,\hat{\gamma}_2,\cdots,\hat{\gamma}_p$ satisfying $\hat{\gamma}_1\geq\hat{\gamma}_2\geq\cdots\geq\hat{\gamma}_p$. The optimal solution of \eqref{eq.lag22}, denoted by $\bY^*$, can be computed as follows:
\begin{enumerate}
  \item when $\hat{\gamma}_p>0$, if $\hat{\gamma}_1/\hat{\gamma}_p\leq \kappa_n$, we have
  \begin{equation*}
    \bY^{*} = \cY;
  \end{equation*}
if $\hat{\gamma}_1/\hat{\gamma}_p > \kappa_n$, we have
  \begin{equation*}
    \bY^{*} = \bU\bGamma^*\bU^{\t},
  \end{equation*}
where $\bGamma^* = \diag(\kappa_n\nu^*,\cdots,\kappa_n\nu^*, \hat{\gamma}_{\alpha+1},\cdots,\hat{\gamma}_{\beta-1},\nu^*,\cdots,\nu^*)$,
$
  \nu^{*} = \frac{\mathop\kappa_{n}\sum_{i=1}^{\alpha^{*}}\hat{\gamma}_i+\sum_{i=\beta^{*}}^{p}\hat{\gamma}_i} {\alpha^{*}\kappa_{n}^2+p-\beta^{*}+1},
$
$\alpha^*$ is the largest index such that $\hat{\gamma}_{\alpha + 1}\geq\kappa_n\nu^*>\hat{\gamma}_{\alpha + 1}$, $\beta^*$ is the smallest index such that $\hat{\gamma}_{\beta - 1}> \nu^*\geq\hat{\gamma}_{\beta}$, and the optimal $\alpha^{*}$ and $\beta^{*}$ can be determined in $\cO(p)$ operations;
  \item when $\hat{\gamma}_p\leq 0$, if $\hat{\gamma}_1\leq 0$, we have
  \begin{equation*}
    \bY^{*} = \mathbf{0};
  \end{equation*}
  if $\hat{\gamma}_1>0$, let $s$ be the largest index such that $\hat{\gamma}_s>0$, we have
  \begin{equation*}
    \bY^{*} = \bU\bGamma^*\bU^{\t},
  \end{equation*}
   where $\bGamma^* = \diag(\kappa_n\nu^*,\cdots,\kappa_n\nu^*, \hat{\gamma}_{\alpha+1},\cdots,\hat{\gamma}_{\beta-1},\nu^*,\cdots,\nu^*)$,
$
  \nu^{*} = \frac{\mathop\kappa_{n}\sum_{i=1}^{\alpha^{*}}\hat{\gamma}_i+\sum_{i=\beta^{*}}^{s}\hat{\gamma}_i} {\alpha^{*}\kappa_{n}^2+p-\beta^{*}+1},
$
$\alpha^*$ is the largest index such that $\hat{\gamma}_{\alpha + 1}\geq\kappa_n\nu^*>\hat{\gamma}_{\alpha + 1}$, $\beta^*$ is the smallest index such that $\hat{\gamma}_{\beta - 1}> \nu^*\geq\hat{\gamma}_{\beta}$, and the optimal $\alpha^{*}$ and $\beta^{*}$ can be determined in $\cO(s)$ operations.
\end{enumerate}
\end{proposition}
\begin{proof}
The result is just a summarization of the Lemma~2.2 in \citep{LiLV20} and Theorem~3.1 in \citep{Wang21}, so we omit it here.
\end{proof}
\begin{remark}
\rm
According to Proposition~\ref{Prop1}, the optimization problem \eqref{eq.lag22} is to find the projection onto the positive semidefinite cone with a specified condition number, and performs truncation of the extreme eigenvalues to control the ill-conditioning of the estimates. This can also be achieved by \eqref{eq.opt3} but requiring two tuning parameters $\tau_1$ and $\tau_2$. The main computational complexity of updating $\bY$ lies in calculating the spectral decomposition, which is also required  in the eigenvalue constrained models. Thus we can find that in one loop of iteration the proposed method has comparable computational burden with the eigenvalue constrained estimators but has better control on the numerical stability of the covariance matrix estimator.
\end{remark}

To update $\bSigma$, the optimization problem \eqref{eq.lag1} can be written as
\begin{equation}
\label{eq.lag11}
  \bSigma^{k+1} = \mathop\argmin_{\bSigma} \frac{1}{2} \left\|\bSigma - \frac{\mu}{1+\mu}\cZ\right\|_F^2 + \frac{\mu\lambda}{1+\mu}\left\|\bSigma\right\|_{\onf}
\end{equation}
with $\cZ = \wbSigma_n - \bLambda^{k} + \frac{1}{\mu}\bY^{k+1}$. The optimization problem \eqref{eq.lag11} has the following closed-form solution and the elements of $\bSigma^{k+1}$ are given by
\begin{equation}
\label{eq.threshold}
  \bSigma_{ij}^{k+1} = \frac{\mu}{1+\mu}\sign\left(\cZ_{ij}\right)\left(\left|\cZ_{ij}\right|-\lambda\right)_{+}.
\end{equation}
This explicit solution makes the updating of $\bSigma$ computationally efficient. Then we can summarize the iterative procedure \eqref{eq.lags} as the Algorithm~\ref{Alg1}.
\begin{algorithm}[htbp]
{\small\caption{Alternating direction algorithm for RWS estimator.}\label{Alg1}
{\textbf{Input:}  $\mu$, $\bSigma^0$ and $\bLambda^0$ .}\\
{\textbf{Output:} optimal approximation $\hat{\Sigma}$.}
\begin{enumerate}
  \item update $\bY$ by solving \eqref{eq.lag22};
  \item update $\bSigma$ by \eqref{eq.threshold};
  \item update $\bLambda$ by $\bLambda^{k+1} = \bLambda^{k} + \frac{1}{\mu}\left(\bSigma^{k+1} - \bY^{k+1}\right)$.
\end{enumerate}}
\end{algorithm}

Since the optimization problem \eqref{eq.propmodel} is convex, we can also prove that the proposed iterative procedure \eqref{eq.lags} has a global convergence property similar to the results given by \citep{XueMZ12,LiuWZ14}.
\begin{theorem}
\label{Thm2}
Given any starting point $\left(\bSigma^0, \bY^0, \bLambda^0\right)$, the sequence $\left\{\left(\bSigma^{k}, \bY^{k}, \bLambda^{k}\right)\right\}$ produced by the proposed iterative procedure \eqref{eq.lags} converges to the optimal solution of \eqref{eq.eqvamodel}.
\end{theorem}
\begin{proof}
See Appendix~\ref{AppendD}.
\end{proof}

In the practical implementation of Algorithm~\ref{Alg1}, we need to specify some details. To stop the iteration procedure, we use the following criterion given by \cite{LiuWZ14}
\begin{eqnarray*}
  \max\left\{\frac{\|\bSigma^{k+1} - \bSigma^{k}\|_F^2}{\|\bSigma^{k}\|_F^2}, \frac{\|\bSigma^{k+1} - \bY^{k+1}\|_F^2}{\|\bSigma^{k}\|_F^2}\right\}<\epsilon,
\end{eqnarray*}
where $\epsilon >0$ is tolerance parameter and we set $\epsilon = 10^{-6}$ in the numerical experiment. Though the penalty parameter $\mu$ has little influence on the theoretical convergence property of Algorithm~\ref{Alg1}, it does have non-negligible impact on the practical numerical performance \citep{Boyds11,CuiLS16}. To improve efficiency, we can search $\mu$ over the interval $[0.1, 10]$ with a step size of 0.5 and select the optimal value. Since the primary goal of this paper is to address the limitations of RATE, if RATE is well-conditioned, it is directly used as the final estimate; otherwise, RATE can be used as the initial value for $\bSigma^0$ and $\bY^0$. Moreover, the initial value of Lagrangian multiplier matrix $\bLambda$ is set to be $\mathbf{0}$.

\subsection{Extensions}
We need to point out that the idea in this paper can be easily extended to more complex covariance matrix estimation models to improve the numerical stability of the estimators. Here, we just present some straightforward generalizations.

Considering the relationship between covariance matrix and correlation matrix, one may first estimate the correlation matrix and then construct the final robust covariance matrix estimator. To estimate a robust and well-conditioned correlation matrix, we may use the following model
\begin{eqnarray}\label{eq.cor}
  \widehat{\bR} = \mathop\argmin_{\bR \in \cP_{+}} \frac{1}{2}\left\|\bR - \widetilde{\bR}_n\right\|_F^2 + \lambda\left\|\bR\right\|_{\onf},\textrm{ s.t. }
\left\{
  \begin{array}{ll}
    \bR_{ij} = b_{ij}, & \hbox{$(i,j)\in \Omega$,} \\
    \mathrm{Cond}\left(\bR\right)\leq \kappa_n, & \hbox{}
  \end{array}
\right.
\end{eqnarray}
where $\widetilde{\bR}_n$ is some robust pairwise correlation matrix estimate, and $\Omega$ is the index set that can be used to constrain $\widehat{\bR}$ having specific structure and of certain interest \citep{High02}. The structure constraint $\left\{\bR_{ij} = b_{ij}|(i,j)\in \Omega\right\}$ does not change the convexity of \eqref{eq.cor}, and should be limited to $\left\{\bR_{ii} = 1| i \in \{1,\cdots,p\}\right\}$ in correlation matrix estimation. Similar to the iterative procedure \eqref{eq.lags} in Section~\ref{sec:algorithm} and with some symbol substitutions,  the optimization problem \eqref{eq.cor} can also be easily solved and the only thing we need to modify is the updating of $\bR$ that is given by
\begin{eqnarray*}
\label{eq.thresholdR}
  \bR_{ij}^{k+1} =
\left\{
     \begin{array}{ll}
     b_{ij}, & \hbox{$(i,j)\in \Omega$,} \\
     \frac{\mu}{1+\mu}\sign\left(\cZ_{ij}\right)\left(\left|\cZ_{ij}\right|-\lambda\right)_{+}, & \hbox{$(i,j)\notin \Omega$,}
  \end{array}
\right.
\end{eqnarray*}
where $\cZ = \widetilde{\bR}_n - \bLambda^{k} + \frac{1}{\mu}\bY^{k+1}$. Note that the rank-based estimator can be readily used to give the RWS estimation of high-dimensional correlation matrix. Moreover, if we set $\widetilde{\bR}_n$ to be the sample correlation matrix, model \eqref{eq.cor} reduces to the well-conditioned version of the estimators proposed by \cite{LiuWZ14} and \cite{CuiLS16}.

The $L_1$ regularization in \eqref{eq.propmodel} does not consider the heteroscedasticity in covariance matrix estimation and may produce biased estimates. To reduce the estimation bias, the adaptive strategy and nonconvex penalty may be used. We present it in the following generic form
\begin{eqnarray}
\label{eq.penalG}
\hat{\bSigma} &=& \mathop\argmin_{\bSigma \in \cP_{+}} \frac{1}{2}\left\|\bSigma - \wbSigma_n\right\|^2_F + P_{\lambda}(\bW,\bSigma), \quad\textrm{s.t.}\;
\mathrm{Cond}\left(\bSigma\right)\leq \kappa_n,
\end{eqnarray}
where $\bW$ is the weight matrix and $P_{\lambda}(\cdot,\cdot)$ is the penalty function. If we just set penalty function $P_{\lambda}(\bW,\bSigma) = \lambda\left\|\bW\odot\bSigma\right\|_{\onf}$, then we can get the adaptive RWS estimator
\begin{eqnarray}
\label{eq.adpRWS}
\hat{\bSigma} &=& \mathop\argmin_{\bSigma\in \cP_{+}} \frac{1}{2}\left\|\bSigma - \wbSigma_n\right\|^2_F + \lambda\left\|\bW\odot\bSigma\right\|_{\onf}, \quad\textrm{s.t.}\;
\mathrm{Cond}\left(\bSigma\right)\leq \kappa_n,
\end{eqnarray}
and the optimization problem \eqref{eq.adpRWS} is still convex and can also be easily solved with the iterative procedure \eqref{eq.lags}. We just need to change the updating of the elements of $\bSigma$ by
\begin{equation}
\label{eq.thresW}
  \bSigma_{ij}^{k+1} = \frac{\mu}{1+\mu}\sign\left(\cZ_{ij}\right)\left(\left|\cZ_{ij}\right|-\lambda\bW_{ij}\right)_{+}.
\end{equation}
However, when $P_{\lambda}(\bW,\bSigma)$ is some nonconvex penalty function, the optimization problem \eqref{eq.penalG} may be nonconvex and only has local minimizers. But this does not mean our idea will not be applicable. A easy check of the algorithms used in \citep{LiuWZ14,WangKW24} will show that our idea can be readily incorporated with almost no additional computational burden in one loop of iteration.

\section{Numerical experiment}
In this section, we present numerical examples to demonstrate the efficiency of the proposed estimator and its superiority over existing methods, using both synthetic and real-world datasets.

\subsection{Selection of tuning parameters}
In the proposed RWS estimator, we need to specify two tuning parameters $\lambda$ and $\kappa_n$ that used to control the sparsity and condition number of the estimates respectively. Here we propose the following grid-search $N$-fold cross-validation strategy to choose these parameters.

Specifically, we randomly split the dataset $N$ times, and in each time the dataset are separated into two parts with training sample size $n_1 = 3/4n$ and testing sample size $n_2 = n - n_1$. The tuning parameters $\lambda$ and $\kappa_n$ are chosen by minimizing
\begin{eqnarray*}
  (\hat{\lambda},\hat{\kappa}_n) = \mathop\argmin_{\lambda>0, \kappa_n\geq 1} \sum_{i=1}^N\left\|\widehat{\bSigma}_{\lambda, \kappa_n}^{train} - \wbSigma_n^{test} \right\|_F^2,
\end{eqnarray*}
over a two dimension grid $(\lambda,\kappa_n)\in [\lambda_0, \lambda_1]\times[\kappa_{n0}, \kappa_{n1}]$, where $\widehat{\bSigma}_{\lambda,\kappa_n}^{train}$ is the RWS estimate with given $\lambda$ and $\kappa_n$ on the training dataset, $\wbSigma_n^{test}$ is the pilot estimate obtained from the testing dataset, $[\lambda_0, \lambda_1]$ is the candidate interval for searching optimal $\hat{\lambda}$, and  $[\kappa_{n0}, \kappa_{n1}]$ is for $\hat{\kappa}_n$.

The above criterion is also suitable for the positive definite high-dimension covariance\textbackslash correlation matrix estimation, in which the lower bound of the eigenvalue and sparsity parameter need to specified \citep{XueMZ12,LiuWZ14,CuiLS16,WangKW24}. But, in practical implementation, the grid-search $N$-fold cross-validation may be time-consuming. It is more practical to use predefined values, such as setting the lower bound $\tau$ as done in \citep{CuiLS16,XueMZ12}.
According to the condition number theory \citep[Ch.~15]{High02b}, we only need to choose an appropriate $\kappa_n$ to ensure that the estimation is not too ill-conditioned, without requiring an exact estimate of its value.
Therefore, in practice, we set $\kappa_n$ to large values corresponding to $\log_{10} \kappa_n\in\{4,5,6\}$, which proves sufficient to ensure numerical stability.

\subsection{Synthetic data generation}
To generate the synthetic data, we set the underlying true covariance matrix has the following two different covariance structures
\begin{enumerate}
  \item Banded matrix: $\bSigma^* = [\sigma^*_{ij}]$ with $\sigma^*_{ij} = \max(1 - (i-j)/10, 0)$.
  \item Block diagonal matrix: $\bSigma^* = \mathrm{blockdiag}(\bSigma_1^*, \bSigma_2^*)$, where $\bSigma_1^* = \bA + \delta \bI_{p/2}$ with independent $a_{ij} = \mathrm{Unif}(0.3,0.8)\times \mathrm{Binorm}(1,0.2)$ and $\delta = \max\{-\lambda_{\min}(A),0\} + 0.001$ to ensure $\bSigma_1^*$ is positive definite, $\bSigma_2^* = 4\bI_{p/2}$.
\end{enumerate}
For each covariance structure, three different scenarios for the distribution of $X$ will be used: (1) Normal: the multivariate normal distribution with zero mean vector; (2) T(3.5): the multivariate $t$ distribution with 3.5 degrees; (3) ST(4): the skewed multivariate $t$ distribution with 4 degrees of freedom and skew parameter equal to 10; (4) CT(5): the contaminated $t$ distribution defined by $X = (1 - b)X_1 + bX_2$, where $X_1$ is from multivariate $t$ distribution with 5 degrees, $X_2$ is from  $\cN(-5, \bI_p)$, and $b$ is from $\mathrm{Binorm}(1, 0.1)$.

\begin{example}
\label{example1}
\rm
In this example, we will give a comparison of the covariance matrix estimators and the following methods will be considered. (1) $\mathrm{SAM}$: the sample covariance matrix;  (2) $\mathrm{RATE}$: the RATE proposed by \cite{AvelBFL18}; (3) $\mathrm{RPDE}$: the robust PDE which is a robust version of \eqref{eq.opt2} with $\widehat{\bSigma}$ replaced by the pilot estimator $\wbSigma_n$; (4) $\mathrm{RWS}$: the proposed RWS estimator \eqref{eq.propmodel}; (5) $\mathrm{ARWS_{1}}$: the adaptive RWS estimator \eqref{eq.adpRWS} with weighted matrix equal to pilot estimator $\wbSigma_n$; (6) $\mathrm{ARWS_{2}}$: the adaptive RWS estimator \eqref{eq.adpRWS} with weighted matrix $\bW = [(({\wbSigma_{n}^{\cT}})_{ij} + 1/n)^{-1}]$. To compare the performance of the estimators, we use Spectral (Spec) and Frobenius (Frob) norms to assess the accuracy. To measure sparse structure identifiability, we report the false selection loss (FSL) computed  as $(\mathrm{FP+FN})/p^2$ in percentage, where $\mathrm{FP}$ is false positive and $\mathrm{FN}$ is false negative. The percentages of the positive definite (PD) estimates in repeated simulations are also reported.

In our experiment, we set the sample size $n = 100$ and $p = 100, 200, 300$. Under this setting, the minimum eigenvalues ($\lambda_{\min}$) and the condition numbers ($\mathrm{Cond}$) of the true covariance matrices is given in Table~\ref{Table1}.
\begin{table}[htp]
 \footnotesize
  \centering
  \caption{Minimum eigenvalue and condition number of true covariance matrices.}\label{Table1}
\begin{tabular}{|l|c|c|c|c|c|c|}
\hline
    & \multicolumn{2}{c|}{$p=100$} & \multicolumn{2}{c|}{$p=200$} & \multicolumn{2}{c|}{$p=300$} \\
\hline
    & $\lambda_{\min}$ & $\mathrm{Cond}$ & $\lambda_{\min}$ & $\mathrm{Cond}$  & $\lambda_{\min}$ & $\mathrm{Cond}$  \\
\hline
Banded         & 2.1$\times10^{-3}$ & 4.8$\times10^{3}$ & 5.6$\times10^{-4}$  & 1.7$\times10^{4}$  & 2.5$\times10^{-4}$  &  3.8$\times10^{4}$  \\
\hline
Block Diagonal & 1.0$\times10^{-3}$ & 7.9$\times10^{3}$ & 1.0$\times10^{-3}$ & 1.6$\times10^{4}$  & 1.0$\times10^{-3}$ & 2.2$\times10^{4}$ \\
\hline
\end{tabular}
\end{table}
To estimate the covariance matrix, we need to specify some details. For the $\mathrm{RATE}$ method, we follow the strategies given by \cite{AvelBFL18} to choose the tuning parameters and only report the results of Huber estimator for its simplicity and numerical performance. The Huber estimator is also used as the pilot estimator in the RPDE and RWS methods. For the RPDE method, we set the lower bound of the eigenvalue $\tau = 10^{-4}$ as \citet{XueMZ12}. For the RWS methods, we may select $\kappa_n$ based on orders of magnitude and the candidates set can be given by $\{10^3,10^4,10^5\}$. For the sparsity tuning parameter $\lambda$ in these estimators, we select it from the interval $[0.01,1]$ with a step size of 0.05. Besides, when the data is from normal distribution, the sample covariance matrix will be used as the pilot estimator.
\begin{table}[htp]
\footnotesize
\caption{Comparison of covariance matrix estimators for banded structure.}\label{Table2}
\centering
\begin{tabular}{lllrrrrrr}
\hline
& $p$  &  & $\mathrm{SAM}$ & $\mathrm{RATE}$ & $\mathrm{RPDE}$ & $\mathrm{RWS}$ & $\mathrm{ARWS_{1}}$ & $\mathrm{ARWS_{2}}$ \\
\hline
\multirow{12}{*}{Normal}&\multirow{4}{*}{$100$}
     &Spec &  5.20(0.78) & 4.36(0.50) & 3.54(0.48) & 3.49(0.50) & 3.29(0.53) & 3.11(0.56)\\
  &  &Frob & 10.23(0.72) & 9.47(0.75) & 7.15(0.71) & 7.09(0.71) & 6.91(0.68) & 6.81(0.74)\\
  &  &FSL  & 0.81(0.00)  & 0.11(0.04) & 0.28(0.09) & 0.28(0.08) & 0.49(0.06) & 0.51(0.02)\\
  &  &PD   & 44   & 0  & 100  & 100  & 100  & 100\\
\cline{2-9}
  &\multirow{4}{*}{$200$}
     &Spec & 8.55(1.02) & 5.06(0.47) & 4.21(0.37) & 4.17(0.38) & 4.05(0.51) & 3.77(0.43)\\
  &  &Frob &20.41(0.82) &14.59(0.77) &11.42(0.60) &11.35(0.59) &10.75(0.79) &10.80(0.70)\\
  &  &FSL  & 0.90(0.00) & 0.06(0.01) & 0.23(0.04) & 0.23(0.05) & 0.40(0.04) & 0.46(0.02)\\
  &  &PD   &  0  &  0 &  100 & 100  &  100 & 100 \\
\cline{2-9}
  &\multirow{4}{*}{$300$}
     &Spec & 11.40(1.13) & 5.31(0.43) & 4.53(0.35) & 4.48(0.35) & 4.17(0.45) & 4.05(0.53)\\
  &  &Frob & 30.50(0.93) &18.75(0.72) &14.91(0.68) &14.77(0.69) &13.34(0.75) &13.98(0.92)\\
  &  &FSL  &  0.93(0.00) & 0.04(0.01) & 0.15(0.06) & 0.14(0.05) & 0.35(0.02) & 0.40(0.02)\\
  &  &PD   &  0 & 0 & 100 & 100 & 100 & 100\\
\hline
\multirow{12}{*}{T(3.5)}&\multirow{4}{*}{$100$}
     &Spec & 42.57(30.02) & 5.43(0.94) & 5.88(1.18) &  5.88(1.18) & 6.96(1.33) & 5.68(1.05) \\
  &  &Frob & 57.07(30.08) &11.88(1.43) &15.59(3.03) & 15.59(3.03) &18.28(3.00) &17.24(2.19) \\
  &  &FSL  & 0.81(0.00)   & 0.57(0.15) & 0.08(0.02) &  0.08(0.02) & 0.10(0.04) & 0.26(0.15) \\
  &  &PD   &  45  & 42  & 100  & 100  & 100  & 100 \\
\cline{2-9}
  &\multirow{4}{*}{$200$}
     &Spec & 101.93(150.16) & 6.93(0.99) & 6.72(1.44) & 6.72(1.44) & 8.58(2.34) & 6.05(1.22)\\
  &  &Frob & 128.81(146.47) &20.72(2.02) &22.27(3.96) &22.27(3.96) &26.54(4.73) &24.40(2.94)\\
  &  &FSL  & 0.91(0.00)     & 0.37(0.19) & 0.05(0.01) & 0.05(0.01) & 0.11(0.10) & 0.19(0.11)\\
  &  &PD   &  0 & 0  & 100  & 100  & 100  & 100 \\
\cline{2-9}
  &\multirow{4}{*}{$300$}
     &Spec & 126.92(158.95) & 7.62(1.02) & 7.24(1.28) & 7.24(1.28) & 9.26(2.24) & 6.21(1.01)\\
  &  &Frob & 167.18(154.21) &28.08(2.65) &27.19(4.08) &27.19(4.08) &32.35(5.26) &29.76(2.73)\\
  &  &FSL  &   0.93(0.00)   & 0.23(0.15) & 0.04(0.01) & 0.04(0.01) & 0.11(0.09) & 0.14(0.09)\\
  &  &PD   &  0  &  0 & 100  & 100  & 100  & 100 \\
\hline
\multirow{12}{*}{ST(4)}&\multirow{4}{*}{$100$}
     &Spec & 7.94(0.23) & 8.58(0.14) & 7.96(0.20) & 7.96(0.20) & 8.67(0.17) & 8.62(0.17)\\
  &  &Frob &19.31(0.43) &21.58(0.25) &19.65(0.43) &19.65(0.43) &21.32(0.52) &21.22(0.51)\\
  &  &FSL  & 0.81(0.00) & 0.75(0.08) & 0.57(0.02) & 0.57(0.02) & 0.06(0.00) & 0.06(0.00)\\
  &  &PD   & 41 & 95 &  100 & 100 & 100 & 100\\
\cline{2-9}
  &\multirow{4}{*}{$200$}
     &Spec & 8.49(1.01) & 8.83(0.11) & 8.31(0.14) & 8.31(0.14) & 8.85(0.14) & 8.81(0.14)\\
  &  &Frob &27.90(0.61) &30.78(0.22) &28.24(0.38) &28.24(0.38) &30.54(0.47) &30.40(0.46)\\
  &  &FSL  & 0.90(0.00) & 0.84(0.08) & 0.64(0.01) & 0.64(0.01) & 0.04(0.00) & 0.03(0.00)\\
  &  &PD   & 0  & 91  & 100  & 100 & 100 &  100\\
\cline{2-9}
  &\multirow{4}{*}{$300$}
     &Spec & 8.91(2.52) & 8.97(0.08) & 8.49(0.12) & 8.49(0.12) & 8.88(0.11) & 8.84(0.11)\\
  &  &Frob &34.45(1.37) &37.75(0.23) &34.67(0.38) &34.68(0.38) &37.30(0.49) &37.12(0.49)\\
  &  &FSL  & 0.93(0.00) & 0.89(0.06) & 0.68(0.01) & 0.68(0.01) & 0.03(0.00) & 0.02(0.00)\\
  &  &PD   & 0 & 84 & 100 & 100 & 100 & 100\\
\hline
\multirow{12}{*}{CT(5)}&\multirow{4}{*}{$100$}
     &Spec & 217.43(66.70) & 8.62(0.14) & 7.01(0.61) & 7.00(0.61) & 7.00(0.61) & 7.00(0.61)\\
  &  &Frob & 218.77(66.15) &23.98(0.30) &31.21(4.15) &31.21(4.15) &31.21(4.15) &31.21(4.15)\\
  &  &FSL  &   0.81(0.00)  & 0.18(0.00) & 0.17(0.00) & 0.17(0.00) & 0.17(0.00) & 0.17(0.00)\\
  &  &PD   & 56 & 100 & 100 & 100 & 100 & 100\\
\cline{2-9}
  &\multirow{4}{*}{$200$}
     &Spec & 476.66(121.58) & 8.65(0.12) & 6.91(0.57) & 6.91(0.57) & 6.91(0.57) & 6.91(0.57)\\
  &  &Frob & 478.16(121.24) &34.21(0.35) &45.72(5.53) &45.72(5.53) &45.72(5.53) &45.72(5.53)\\
  &  &FSL  & 0.90(0.00)     & 0.08(0.00) & 0.08(0.00) & 0.08(0.00) & 0.08(0.00) & 0.08(0.00)\\
  &  &PD   & 0    & 100 & 100 & 100 & 100 & 100\\
\cline{2-9}
  &\multirow{4}{*}{$300$}
     &Spec & 667.15(202.21) & 8.69(0.14) & 7.11(0.61) & 7.11(0.61) & 7.11(0.61) & 7.11(0.61)\\
  &  &Frob & 669.26(201.51) &41.93(0.51) &54.42(7.23) &54.42(7.23) &54.42(7.23) &54.42(7.23)\\
  &  &FSL  &   0.93(0.00)   & 0.05(0.00) & 0.05(0.00) & 0.05(0.00) & 0.05(0.00) & 0.05(0.00)\\
  &  &PD   &  0  & 100 & 100 & 100 & 100 & 100\\
\hline
\end{tabular}
\end{table}

\begin{table}[htp]
\footnotesize
\caption{Comparison of covariance matrix estimators for block diagonal structure.}\label{Table3}
\centering
\begin{tabular}{lllrrrrrr}
\hline
& $p$  &  & $\mathrm{SAM}$ & $\mathrm{RATE}$ & $\mathrm{RPDE}$ & $\mathrm{RWS}$ & $\mathrm{ARWS_{1}}$ & $\mathrm{ARWS_{2}}$ \\
\hline
\multirow{12}{*}{Normal}&\multirow{4}{*}{$100$}
     &Spec & 8.17(0.64) & 5.68(0.13) & 5.67(0.07) & 5.61(0.08) & 5.61(0.08) & 5.61(0.08)\\
  &  &Frob &26.03(0.51) &13.45(0.19) &12.53(0.07) &12.57(0.08) &12.57(0.08) &12.57(0.08)\\
  &  &FSL  & 0.94(0.00) & 0.05(0.01) & 0.04(0.00) & 0.04(0.00) & 0.04(0.00) & 0.04(0.00)\\
  &  &PD   & 50 & 100 & 100 & 100 & 100 & 100\\
\cline{2-9}
  &\multirow{4}{*}{$200$}
     &Spec & 17.26(0.91) & 11.55(0.11) & 11.64(0.06) & 11.50(0.07) &11.50(0.07) &11.50(0.07)\\
  &  &Frob & 64.77(0.85) & 27.60(0.29) & 26.02(0.08) & 26.12(0.10) &26.12(0.10) &26.12(0.10)\\
  &  &FSL  &  0.94(0.00) &  0.05(0.00) &  0.04(0.00) &  0.04(0.00) & 0.04(0.00) & 0.04(0.00)\\
  &  &PD   & 0  & 100  & 100  & 100  & 100  & 100 \\
\cline{2-9}
  &\multirow{4}{*}{$300$}
     &Spec & 27.65(1.52) & 16.96(0.10) & 17.15(0.06) & 16.87(0.07) & 16.87(0.07) & 16.87(0.07)\\
  &  &Frob &113.10(1.20) & 41.30(0.33) & 39.05(0.12) & 39.19(0.14) & 39.19(0.14) & 39.19(0.14)\\
  &  &FSL  & 0.94(0.00)  &  0.04(0.00) &  0.04(0.00) &  0.04(0.00) &  0.04(0.00) & 0.04(0.00)\\
  &  &PD   & 0 & 100  & 100 & 100 & 100 & 100\\
\hline
\multirow{12}{*}{T(3.5)}&\multirow{4}{*}{$100$}
     &Spec & 99.82(91.26) & 5.38(0.40) & 5.34(0.44) & 5.34(0.44) & 5.34(0.44) & 5.34(0.44)\\
  &  &Frob &136.88(91.71) &16.81(1.80) &22.05(3.03) &22.05(3.03) &22.05(3.03) &22.05(3.03)\\
  &  &FSL  &  0.93(0.00)  & 0.05(0.01) & 0.05(0.00) & 0.05(0.00) & 0.05(0.00) & 0.05(0.00)\\
  &  &PD   &  53  & 100  & 100  & 100  &  100 & 100 \\
\cline{2-9}
  &\multirow{4}{*}{$200$}
     &Spec & 310.66(742.00) & 10.35(0.43) & 9.40(0.45) & 9.40(0.45) & 9.40(0.45) & 9.40(0.45)\\
  &  &Frob & 394.63(727.74) & 31.30(2.50) &36.18(3.79) &36.18(3.79) &36.18(3.79) &36.18(3.79) \\
  &  &FSL  &   0.94(0.00)   &  0.04(0.00) & 0.04(0.00) & 0.04(0.00) & 0.04(0.00) & 0.04(0.00)\\
  &  &PD   &  0  & 100 & 100 & 100 & 100 & 100\\
\cline{2-9}
  &\multirow{4}{*}{$300$}
     &Spec & 448.40(846.37) & 15.61(0.53) & 14.92(0.51) & 14.92(0.51) & 14.92(0.51) & 14.92(0.51)\\
  &  &Frob & 594.61(826.69) & 46.38(3.57) & 49.01(4.46) & 49.01(4.46) & 49.01(4.46) & 49.01(4.46)\\
  &  &FSL  &   0.94(0.00)   &  0.04(0.00) &  0.04(0.00) &  0.04(0.00) &  0.04(0.00) & 0.04(0.00)\\
  &  &PD   & 0 & 100 & 100 & 100 & 100 & 100\\
\hline
\multirow{12}{*}{ST(4)}&\multirow{4}{*}{$100$}
     &Spec & 7.32(2.71) & 7.68(0.12) & 7.22(0.18) & 7.22(0.18) & 7.25(0.18) & 7.23(0.18)\\
  &  &Frob &23.22(1.67) &25.31(0.12) &23.02(0.25) &23.02(0.25) &23.03(0.26) &23.03(0.26)\\
  &  &FSL  & 0.94(0.00) & 0.91(0.01) & 0.28(0.02) & 0.28(0.02) & 0.11(0.01) & 0.13(0.01)\\
  &  &PD   & 46 & 100  & 100  & 100  & 100  & 100 \\
\cline{2-9}
  &\multirow{4}{*}{$200$}
     &Spec &15.25(25.92) & 14.01(0.21) & 13.24(0.31) & 13.24(0.31) & 13.24(0.31) & 13.22(0.30)\\
  &  &Frob &46.98(23.08) & 47.78(0.17) & 44.22(0.29) & 44.22(0.29) & 44.22(0.30) & 44.21(0.29)\\
  &  &FSL  & 0.94( 0.00) &  0.91(0.00) &  0.35(0.01) &  0.35(0.01) &  0.18(0.02)&  0.20(0.01)\\
  &  &PD   &  0 & 99 & 100 & 100 & 100 & 100\\
\cline{2-9}
  &\multirow{4}{*}{$300$}
     &Spec & 18.31(3.13) & 19.63(0.37) & 19.26(0.36) & 19.26(0.36) & 18.69(0.47) & 18.67(0.48)\\
  &  &Frob & 68.02(1.67) & 70.66(0.24) & 66.08(0.36) & 66.08(0.36) & 66.16(0.32) & 66.19(0.32)\\
  &  &FSL  &  0.94(0.00) &  0.91(0.00) &  0.19(0.02) &  0.19(0.02) &  0.26(0.02) &  0.27(0.02)\\
  &  &PD   & 0 & 89 & 100 & 100 & 100 & 100\\
\hline
\multirow{12}{*}{CT(5)}&\multirow{4}{*}{$100$}
     &Spec & 217.43(66.70) & 8.62(0.14) & 7.01(0.61) & 7.00(0.61) & 7.00(0.61) & 7.00(0.61)\\
  &  &Frob & 218.77(66.15) &23.98(0.30) &31.21(4.15) &31.21(4.15) &31.21(4.15) &31.21(4.15)\\
  &  &FSL  &   0.81(0.00)  & 0.18(0.00) & 0.17(0.00) & 0.17(0.00) & 0.17(0.00) & 0.17(0.00)\\
  &  &PD   & 56 & 100 & 100 & 100 & 100 & 100\\
\cline{2-9}
  &\multirow{4}{*}{$200$}
     &Spec & 476.66(121.58) & 8.65(0.12) & 6.91(0.57) & 6.91(0.57) & 6.91(0.57) & 6.91(0.57)\\
  &  &Frob & 478.16(121.24) &34.21(0.35) &45.72(5.53) &45.72(5.53) &45.72(5.53) &45.72(5.53)\\
  &  &FSL  & 0.90(0.00)     & 0.08(0.00) & 0.08(0.00) & 0.08(0.00) & 0.08(0.00) & 0.08(0.00)\\
  &  &PD   & 0    & 100 & 100 & 100 & 100 & 100\\
\cline{2-9}
  &\multirow{4}{*}{$300$}
     &Spec & 667.15(202.21) & 8.69(0.14) & 7.11(0.61) & 7.11(0.61) & 7.11(0.61) & 7.11(0.61)\\
  &  &Frob & 669.26(201.51) &41.93(0.51) &54.42(7.23) &54.42(7.23) &54.42(7.23) &54.42(7.23)\\
  &  &FSL  &   0.93(0.00)   & 0.05(0.00) & 0.05(0.00) & 0.05(0.00) & 0.05(0.00) & 0.05(0.00)\\
  &  &PD   &  0  & 100 & 100 & 100 & 100 & 100\\
\hline
\end{tabular}
\end{table}

For each setting, we repeat the numerical experiment 100 times. The means and standard deviations (in parentheses) of the estimation errors are presented, with all numerical results summarized in Tables ~\ref{Table2} and \ref{Table3}. From these results, it can be clearly observed that the proposed methods effectively enforce positive definite corrections while fully guaranteeing the positive definiteness of the estimation. Specifically, Table~\ref{Table2} demonstrates that the proposed $\mathrm{RWS}$ method and its adaptive variants yield more accurate estimates in terms of both spectral and Frobenius norms, with the $\mathrm{ARWS_{2}}$  method exhibiting particularly superior performance. Regarding sparsity structure identification, under normal distribution data, the $\mathrm{RATE}$ method achieves optimal results but fails to ensure positive definiteness. For heavy-tailed distributions or contaminated data, both $\mathrm{RPDE}$ and $\mathrm{RWS}$ methods demonstrate remarkably similar yet outstanding performance. Table~\ref{Table3} reveals an interesting finding that the $\mathrm{RWS}$ methods produce identical results. This is because the initial value $\bSigma^0$ provided by the $\mathrm{RATE}$ method is not only positive definite but also well-conditioned, and the following condition number correction is not needed.

\end{example}

\begin{example}
\label{example2}
\rm
Example~\ref{example1} demonstrates that the  $\mathrm{PDE}$ and $\mathrm{RWS}$ based methods exhibit comparable performance. This example will show the reason why we prefer the condition number constraint to the eigenvalue constraint through the important application of covariance matrix in discriminant analysis.
Here we only consider the linear discriminant analysis (LDA) for two classes, which has a simple form but suffices to illustrate our purpose.

Suppose the random vector $X$ follows two different multivariate normal distributions, which have different mean vectors but sharing the same covariance matrix
\begin{eqnarray*}
  C_0: X|y = 0 \sim \cN(\bmu_0,\bSigma), \quad  C_1: X|y = 1 \sim \cN(\bmu_1,\bSigma),
\end{eqnarray*}
where $y\in \{0, 1\}$ is the class label. Then, according to the LDA, we can predict the class label by
\begin{eqnarray*}
  f(x) = \mathrm{I}\left\{\ln\left(\frac{\pi_0}{\pi_1}\right) + \left(x - \frac{\bmu_0 + \bmu_1}{2}\right)^T\bSigma^{-1}(\bmu_0 - \bmu_1)<0\right\},
\end{eqnarray*}
where $ \mathrm{I}(\cdot)$ is the indicator function,  $\pi_0$ and $\pi_1$ are the prior probabilities for the two classes. In practical applications, the population parameters $\pi_0$, $\pi_1$, $\bmu_0$ and $\bmu_1$ are estimated by
\begin{eqnarray*}
  \hat{\pi}_0 = \frac{n_0}{n},\; \hat{\pi}_1 = \frac{n_1}{n},\; \hat{\bmu}_{0} = \frac{1}{n_0}\sum_{i=1}^n\bx_i\mathrm{I}(y_i = 0) \;\textrm{ and }\;\hat{\bmu}_{1} = \frac{1}{n_1}\sum_{i=1}^n\bx_i\mathrm{I}(y_i = 1),
\end{eqnarray*}
where $\{\bx_i,y_i\}_{i=1}^n$ is the observed samples,  $n_0$ is the sample size from class $C_0$, and $n_1$ is the sample size from class $C_1$. The covariance matrix $\bSigma$ will be estimated by the PDE and RSC methods, and we will compare their performance with respect to classification accuracy.

To compare the impact of condition number and eigenvalue constraints on classification performance, we vary the magnitudes of the constraints recommended in the literature and record their corresponding classification errors. In the experiment, we fix the dimensionality at $p=200$ and impose a banded covariance structure, where the condition number and minimum eigenvalue of $\bSigma$ are $1.77\times10^{4}$ and $5.62\times10^{-4}$, respectively. For classes $C_0$ and $C_1$, the mean vectors are set to $\mathbf{0}$ and $\mathbf{3}$, respectively. The training dataset consists of 200 observations (100 per class), and the testing dataset also contains 200 observations (100 per class). All datasets are generated from multivariate normal distributions. For each configuration, the experiment is repeated 100 times, and the average classification errors along with its standard deviations (in parentheses)  are reported in Table~\ref{Table4}.
\begin{table}[htp]
\footnotesize
  \centering
  \caption{Classification error comparison between PDE and RWS methods.}\label{Table4}
\begin{tabular}{|c|c|l|l|l|l|}
\hline
\multirow{2}{*}{$\mathrm{PDE}$}
  & $\tau$ &$10^{-3}$   &$10^{-4}$   &$10^{-5}$    &$10^{-6}$   \\
\cline{2-6}
  &  &0.049(0.02) &0.126(0.06) &0.143(0.08)  &0.145(0.08) \\
\hline
\multirow{2}{*}{$\mathrm{RWS}$}
  & $\kappa_n$ &$10^{3}$    &$10^{4}$    &$10^{5}$     &$10^{6}$    \\
\cline{2-6}
  &    &0(0)        &0.048(0.02) &0.126(0.067) &0.143(0.08) \\
\hline
\end{tabular}
\end{table}
From Table~\ref{Table4}, we see that the RWS method outperforms the PDE method, primarily due to its direct control over the condition number, which bounds both the largest and smallest eigenvalues.
Specifically, when the constraint parameters are set too stringently or loosely relative to the underlying true condition number (e.g., the fourth column in Table 4), both methods yield large classification errors, although RWS performs slightly better. However, when the constraints are of similar magnitude to, or slightly over- or under-estimate, the true value (in the first to third columns), the RWS method demonstrates significantly better performance than the PDE method. This difference arises because the PDE method fails to control the largest eigenvalue, often producing ill-conditioned estimates. To illustrate this limitation, we computed the ratios between the condition number of the PDE estimates and that of the true covariance matrix, and present the histograms of these ratios in Figure~\ref{Fig2}.
\begin{figure}[htp]
  \centering
  \includegraphics[width=0.85\textwidth,height=0.6\textwidth]{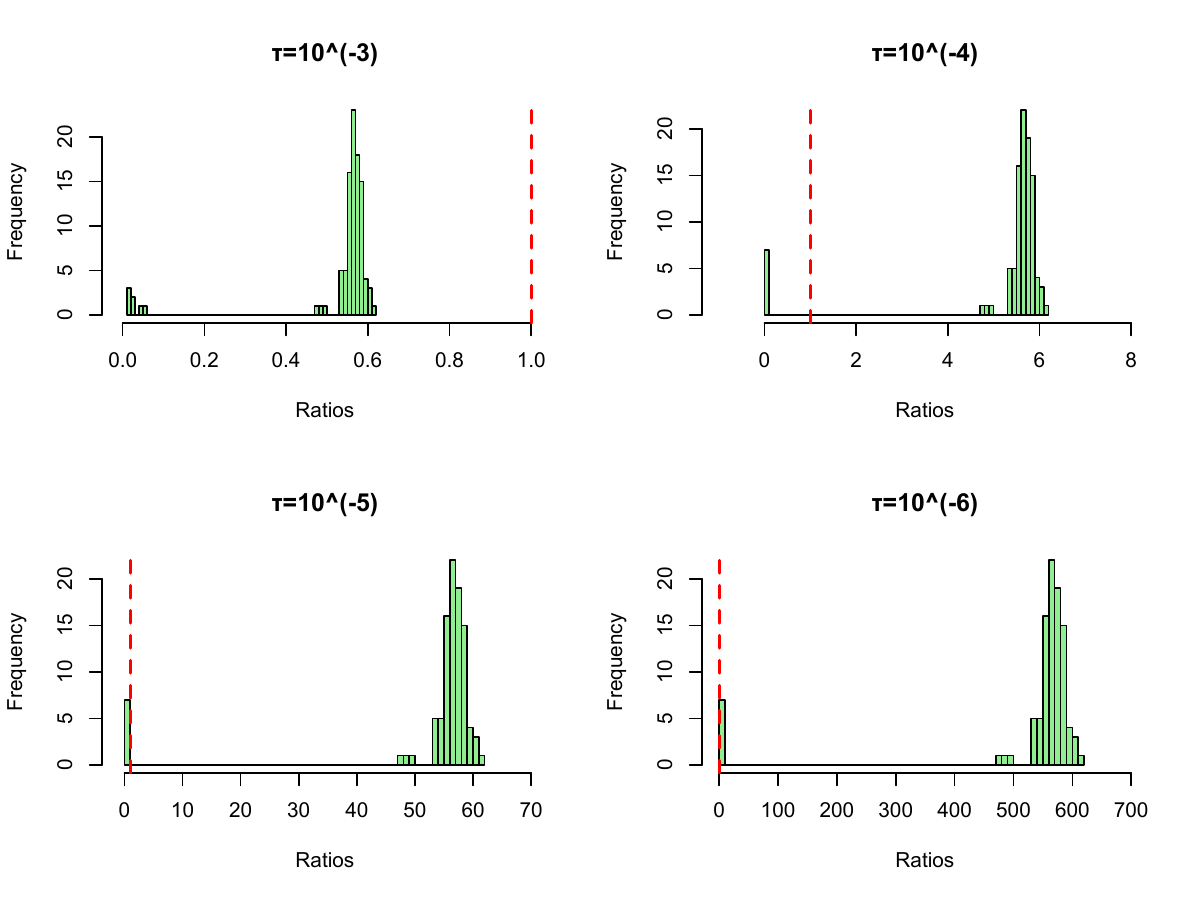}\\
  \caption{The ratios of condition numbers between PDE estimates and true covariance matrix.}\label{Fig2}
\end{figure}
According to Figure~\ref{Fig2}, it is clear that the condition number of the PDE estimates is highly sensitive to the constraint magnitude. Even a slight change in the constraint can lead to a dramatic increase in the condition number. Therefore, we conclude that the condition number constraint (employed by the RWS method) demonstrates clear superiority in ensuring the numerical stability of statistical model computations.
\end{example}

\subsection{Real dataset}
One important application of the covariance matrix is the modern portfolio selection problem in finance, which describes how investors quantify and minimize overall portfolio risk. For a market with $p$ risky assets, the return of asset $i$ over one period, denoted $r_i$, is computed as the change in its closing price over the period divided by its price at the beginning. Then the Markowitz portfolio selection may be formulated as the following quadratic optimization problem
\begin{equation}
\label{eq.Markwz}
 \widehat{\bw} = \mathop\argmin_{\bw\in \mathbb{R}^{p}} \bw^{\t}\bSigma \bw,\quad  \textrm{s.t.}\;\; \bw^{\t}\mathbf{1} = 1,
\end{equation}
where $\bSigma$ is the underlying true covariance matrix of the $p$ risky assets, $\bw$ is the vector of weights for different assets, $\mathbf{1}$ is the vector of all 1s and $\bw^{\t}\mathbf{1} = 1$ describes the budget constraint. In practical applications, $\bSigma$ is unknown and the optimal $\widehat{\bw}$ is usually computed with some empirical version of $\bSigma$, such as the sample covariance matrix. The drawbacks of the sample covariance matrix for high-dimensional portfolio selection have been full addressed by \citet{EKaro10}. From a different view, \citet{WangY21} presents the sensitivity analysis of constrained quadratic optimization problem, which explicitly describes the mechanism of the condition number of $\bSigma$  in controlling the computation error of the optimal weights.

\begin{table}[htp]
 \footnotesize
  \centering
  \caption{Comparison of portfolio selection with different covariance matrix estimators.}\label{Table5}
\begin{tabular}{|l|rrr|rrrr|}
\hline
      & \multicolumn{3}{c|}{RWS}      & \multicolumn{4}{c|}{PDE}         \\
\hline
      &$\kappa_n = 10^{4}$ &$\kappa_n = 10^{5}$& $\kappa_n = 10^{6}$&$\tau = 10^{-4}$&$\tau = 10^{-5}$ & $\tau = 10^{-6}$ & $\tau = 10^{-7}$ \\
\hline
Mean   &  1.691  &  1.711  &  1.711  &  2.141  &  1.277  &  1.711  &  1.711 \\
Std.    &  6.837  &  6.833  &  6.834  &  7.014  &  7.439  &  6.834  &  6.834 \\
Sharp.  & 24.734  & 25.042  & 25.046  & 30.533  & 17.166  & 25.046  & 25.046 \\
\hline
\end{tabular}
\end{table}

To solve the optimization problem \eqref{eq.Markwz}, the empirical estimator of $\bSigma$ should be positive definite. Here, we consider two different empirical covariance matrix estimators: the PDE method and the proposed RWS method. The dataset comprises
monthly return data of companies in the S\&P 100 index ranging from Jenuory 2007 to December 2019.
Stocks with missing values were excluded, and we obtained 88 stocks with 156 observations. We used the fivefold cross-validation to choose the tuning parameters.  To evaluate the effectiveness of covariance matrix estimation methods, we employed a rolling window approach. Using the most recent $n=120$ months of historical return data, the covariance matrix was estimated and used to compute the optimal portfolio weights $\widehat{\bw}$.
These weights $\widehat{\bw}$ were then applied to construct the portfolio for the subsequent month, and its return $u = \widehat{\bw}^{\t}\bx$ was calculated. This procedure was repeated, generating a sequence of 36 out-of-sample portfolio returns. The monthly mean returns, standard deviations (Std.), and the corresponding Sharpe ratios (Sharp.) of these 36 portfolios are expressed in percentage and summarized in Table~\ref{Table5}. According to the numerical results in Table~\ref{Table5}, we can easily find that the RWS method preserves much better stability to the PDE method and exhibits low sensitivity to parameter variations, further validating the effectiveness of the proposed method.

\section{Concluding remark}
In this paper, we propose the RWS covariance matrix estimator to address critical limitations of existing methods in high-dimensional complex data. RWS uniquely integrates a condition number constraint into a robust adaptive thresholding framework, simultaneously ensuring positive definiteness, sparsity, and controlled conditioning, thereby avoiding the indefiniteness and ill-conditioning inherent in pilot estimators and the need for post-hoc corrections. Crucially, RWS uses a single parameter to govern the condition number, simplifying tuning and enhancing numerical stability compared to eigenvalue-bound approaches. Theoretical analysis confirms RWS achieves the minimax optimal convergence rate under the Frobenius norm, and we developed an efficient algorithm with convergence guarantees. Extensive simulations and real-world applications demonstrate RWS's superiority in producing positive definite, well-conditioned, and sparse estimates, enhancing downstream task stability, accuracy, and structural recovery while reducing tuning sensitivity. The idea in this paper is also versatile. It readily extends to well-conditioned correlation matrices and supports adaptive weighting/non-convex penalties for bias mitigation. Our key future directions include developing distributed/online ADM variants for large-scale data, and extending theoretical guarantees to weaker assumptions and precision matrix estimation.

\section*{Data Availability Statement}
The data that support the findings of this study are available from the corresponding author upon reasonable request.

\appendix
\section*{Appendix}
\section{Proof of Lemma~\ref{Lemma1}}
\label{AppendA}
\begin{proof}
The finiteness of the upper bound of condition number $\kappa_n$ makes it sufficient to prove the constraint $\cD = \left\{\bSigma|\Cond\left(\bSigma\right)\leq \kappa_n, \bSigma\in \cP_{++}\right\}$ is convex. Let $\bSigma_1$ and $\bSigma_2$ belong to $\cD$, then according to the definition of $\Cond\left(\bSigma\right)$ we have $\gamma_{\max}(\bSigma_1)\leq \kappa_n \gamma_{\min}(\bSigma_1)$ and $\gamma_{\max}(\bSigma_2)\leq \kappa_n \gamma_{\min}(\bSigma_2)$. For $\forall \alpha \in [0,1]$, we can get
\begin{eqnarray*}
  \Cond\left(\alpha\bSigma_1 + (1-\alpha)\bSigma_2\right) &=& \frac{\gamma_{\max}\left(\alpha\bSigma_1 + (1-\alpha)\bSigma_2\right)}{\gamma_{\min}\left(\alpha\bSigma_1 + (1-\alpha)\bSigma_2\right)} \\
   &\leq & \frac{\gamma_{\max}\left(\alpha\bSigma_1\right) + \gamma_{\max}\left((1-\alpha)\bSigma_2\right)}{\gamma_{\min}\left(\alpha\bSigma_1\right) + \gamma_{\min}\left((1-\alpha)\bSigma_2\right)} \\
   &\leq& \frac{\kappa_n\left(\gamma_{\min}\left(\alpha\bSigma_1\right) + \gamma_{\min}\left((1-\alpha)\bSigma_2\right)\right)}{\gamma_{\min}\left(\alpha\bSigma_1\right) + \gamma_{\min}\left((1-\alpha)\bSigma_2\right)}\\
& = &\kappa_n
\end{eqnarray*}
and $\alpha\bSigma_1 + (1-\alpha)\bSigma_2\in \cD$, which completes the proof.
\end{proof}

\section{Proof of theorem~\ref{Thm1}}
\label{AppendB}
\begin{proof}
Suppose $\bSigma^*$ is a feasible solution to \eqref{eq.propmodel} and set
\begin{equation}
\label{A.eq1}
\hat{\bDelta} = \mathop\argmin_{\bDelta = \bDelta^{\t}, \Cond(\bSigma^{*} + \bDelta)\leq \kappa_n}F(\bDelta) \equiv \frac{1}{2}\left\|\bSigma^{*} + \bDelta - \wbSigma_n\right\|^2_F + \lambda\left\|\bSigma^{*} + \bDelta\right\|_{\onf},
\end{equation}
then we can easily get $\hat{\bDelta} = \hbSigma - \bSigma^{*}$. Then we can see that
\begin{eqnarray}
\label{A.eq2}
  F(\hat{\bDelta}) - F(\mathbf{0})&=& \frac{1}{2}\left\|\bSigma^{*} + \hat{\bDelta} - \wbSigma_n\right\|^2_F + \lambda\left\|\bSigma^{*} + \hat{\bDelta}\right\|_{\onf} - \frac{1}{2}\left\|\bSigma^{*} - \wbSigma_n\right\|^2_F - \lambda\left\|\bSigma^{*}\right\|_{\onf} \nonumber\\
   &=& \frac{1}{2}\left\|\hat{\bDelta}\right\|^2_F  + \left\langle\hat{\bDelta}, \bSigma^{*} - \wbSigma_n\right\rangle +  \lambda\left\|\bSigma^{*} + \hat{\bDelta}\right\|_{\onf} - \lambda\left\|\bSigma^{*}\right\|_{\onf}\leq 0.
\end{eqnarray}
Considering the sparse pattern of $\bSigma^*$, we define the nonzero element index set of  $\bSigma^*$ as  $\mathrm{Supp}(\bSigma^*) = \left\{(i,j)|\bSigma^*_{ij} \neq 0\right\}$. We also define the following two subspace of $\mathbb{R}^{p\times p}$
\begin{equation*}
  \cE = \left\{\bA\in \mathbb{R}^{p\times p}| \bA_{ij} = 0,\; \forall (i,j)\notin \mathrm{Supp}(\bSigma^*)\right\}\textrm{ and }\cE^{c} = \mathbb{R}^{p\times p}\backslash \cE,
\end{equation*}
which can be used to split a matrix $\bA$ into two parts with respect to sparse pattern of $\bSigma^*$. Specifically, we use $\bA_{\cE}$ to denote the projection of matrix $\bA$ onto subspace $\cE $, and we can easily get $\bA_{\cE^c} = \bA - \bA_{\cE}$. Since $\left\|\cdot\right\|_{\onf}$ is decomposable, we can easily get
\begin{eqnarray}
\label{A.eq3}
  \left\|\bSigma^{*} + \hat{\bDelta}\right\|_{\onf} &=& \left\|\bSigma^{*}_{\cE} + \bSigma^{*}_{\cE^c} + \hat{\bDelta}_{\cE} + \hat{\bDelta}_{\cE^c}\right\|_{\onf} \nonumber\\
   &\geq&\left\|\bSigma^{*}_{\cE}  + \hat{\bDelta}_{\cE^c}\right\|_{\onf} -  \left\|\bSigma^{*}_{\cE^c} + \hat{\bDelta}_{\cE}\right\|_{\onf}\nonumber\\
   &=& \left\|\bSigma^{*}_{\cE}\right\|_{\onf} + \left\|\hat{\bDelta}_{\cE^c}\right\|_{\onf} - \left\|\bSigma^{*}_{\cE^c}\right\|_{\onf} -\left\| \hat{\bDelta}_{\cE}\right\|_{\onf}.
\end{eqnarray}
Note that $\left\|\bSigma^{*}\right\|_{\onf} = \left\|\bSigma^{*}_{\cE}\right\|_{\onf} + \left\|\bSigma^{*}_{\cE^c}\right\|_{\onf}$, then with \eqref{A.eq3} we have
\begin{equation}
\label{A.eq4}
  \left\|\bSigma^{*} + \hat{\bDelta}\right\|_{\onf} - \left\|\bSigma^{*}\right\|_{\onf}\geq \left\|\hat{\bDelta}_{\cE^c}\right\|_{\onf}  -\left\| \hat{\bDelta}_{\cE}\right\|_{\onf} - 2\left\|\bSigma^{*}_{\cE^c}\right\|_{\onf}.
\end{equation}
By the Cauchy-Schwartz inequality, we have
\begin{eqnarray}
\label{A.eq5}
  \left\langle\hat{\bDelta}, \bSigma^{*} - \wbSigma_n\right\rangle &\leq&\left\|\bSigma^{*} - \wbSigma_n\right\|_{\max} \left(\left\|\hat{\bDelta}\right\|_{\onf} + \sum_{i=1}^p|\hat{\bDelta}_{ii}| \right).
\end{eqnarray}
Therefore, with \eqref{A.eq4} and \eqref{A.eq5} and from \eqref{A.eq2} we have
\begin{eqnarray}
\label{A.eq6}
  \frac{1}{2}\left\|\hat{\bDelta}\right\|^2_F &\leq& \left\langle\hat{\bDelta}, \wbSigma_n-\bSigma^{*}\right\rangle -  \lambda\left(\left\|\bSigma^{*} + \hat{\bDelta}\right\|_{\onf} - \left\|\bSigma^{*}\right\|_{\onf}\right) \nonumber\\
   &\leq& \left\|\bSigma^{*} - \wbSigma_n\right\|_{\max} \left(\left\|\hat{\bDelta}\right\|_{\onf} + \sum_{i=1}^p|\hat{\bDelta}_{ii}| \right)\nonumber\\
& & - \lambda\left(\left\|\hat{\bDelta}_{\cE^c}\right\|_{\onf}  -\left\| \hat{\bDelta}_{\cE}\right\|_{\onf} - 2\left\|\bSigma^{*}_{\cE^c}\right\|_{\onf}\right).
\end{eqnarray}
According to the definition of $\cE$, we have $\bSigma^{*}_{\cE^c} = \mathbf{0}$ and the matrix in $\cE$ has at most $s$ nonzero elements. Give the event $\|\bSigma^{*} - \wbSigma_n\|_{\max} \leq \lambda$ and with Lemma~\ref{Lemma2}, we can get
\begin{eqnarray*}
  \left\|\hat{\bDelta}\right\|^2_F &\leq & 2\lambda \left(\left\|\hat{\bDelta}_{\cE^c}\right\|_{\onf}  + \left\| \hat{\bDelta}_{\cE}\right\|_{\onf} + \sum_{i=1}^p|\hat{\bDelta}_{ii}|\right)- 2\lambda\left(\left\|\hat{\bDelta}_{\cE^c}\right\|_{\onf}  -\left\| \hat{\bDelta}_{\cE}\right\|_{\onf}\right)\nonumber\\
&=&  2\lambda \left(2\left\| \hat{\bDelta}_{\cE}\right\|_{\onf} + \sum_{i=1}^p|\hat{\bDelta}_{ii}|\right)\nonumber\\
&\leq&\lambda C_1\sqrt{s + p}\left\|\hat{\bDelta}\right\|_F,
\end{eqnarray*}
where $C_1$ is some positive constant. Then, by taking $\lambda = C_0\sqrt{\log p/n}$, the following inequality
\begin{eqnarray*}
  \left\|\hat{\bDelta}\right\|_F &\leq&C_2\sqrt{\frac{(s + p)\log p}{n}}
\end{eqnarray*}
holds with probability tending to one.
\end{proof}

\section{Proof of theorem~\ref{Thm2}}
\label{AppendD}

\begin{proof}
Suppose the Lagrangian function of \eqref{eq.eqvamodel} is given by
\begin{eqnarray*}
\label{eq.lagrange}
  L\left(\bSigma,\bY,\bLambda\right) &=& \frac{1}{2}\left\|\bSigma - \wbSigma_n\right\|^2_F + \lambda\left\|\bSigma\right\|_{\onf} + \left\langle\bSigma-\bY,\bLambda\right\rangle,\;\;
 \textrm{s.t.}\; \left\{
   \begin{array}{ll}
     \Cond\left(\bY\right)\leq \kappa_n, & \hbox{ }\\
     \bY \in \cP_{+}, & \hbox{ }
   \end{array}\right.\nonumber
\end{eqnarray*}
Let $(\hat{\bSigma}^*,\hat{\bY}^*)$ be an optimal solution of \eqref{eq.eqvamodel} and $\hat{\bLambda}^*$ the associated optimal Lagrangian dual variable, then according to the Karush-Kuhn-Tucher (KKT) condition we have
\begin{eqnarray}
  \frac{1}{\lambda}\left(\wbSigma_n - \hat{\bSigma}^* - \hat{\bLambda}^*\right)_{ij} &\in& \partial|\hat{\bSigma}^*_{ij}|, \quad  i\neq j, \label{B.eq2}\\
  \left(\wbSigma_n - \hat{\bSigma}^* - \hat{\bLambda}^*\right)_{ii}&=& 0, \label{B.eq3}\\
  \hat{\bSigma}^*=\hat{\bY}^*,\;\;
  \Cond(\hat{\bY}^*)\leq \kappa_n, &\;\textrm{and}\;& \hat{\bY}^*\in \cP_{+},
\end{eqnarray}
where $i = 1, 2, \cdots, p$, $j = 1, 2, \cdots, p$, and $\partial$ is to calculate the subdifferential. With the optimality of $\left(\hat{\bSigma}^*,\hat{\bY}^*,\hat{\bLambda}^*\right)$, we also have
\begin{equation}
\label{B.eq5}
  \left\langle\bY-\hat{\bY}^*,\hat{\bLambda}^*\right\rangle\leq 0,\quad \forall\; \bY\in \cP_{+} \textrm{ with } \Cond(\bY)\leq \kappa_n.
\end{equation}
According to the updating of $\bY$,  from \eqref{eq.lag22} we have
\begin{equation}
\label{B.eqUpLmd}
  \left\langle\bY^{k+1} - \cY_n,\bY^{k+1} - \cY_n\right\rangle\leq \left\langle\bY^{k+1} - \cY_n,\bY - \cY_n\right\rangle,\quad \forall\; \bY\in \cP_{+} \textrm{ with } \Cond(\bY)\leq \kappa_n,
\end{equation}
and then the following inequality holds
\begin{equation}
\label{B.eq6}
  \left\langle\bLambda^{k} + \frac{1}{\mu}(\bSigma^{k} - \bY^{k+1}),\bY - \bY^{k+1}\right\rangle \leq 0,\quad \forall\; \bY\in \cP_{+} \textrm{ with } \Cond(\bY)\leq \kappa_n.
\end{equation}
With the following updating formula of $\bLambda$
\begin{equation}
\label{B.eq77}
  \bLambda^{k+1} = \bLambda^{k} + \frac{1}{\mu}\left(\bSigma^{k+1} - \bY^{k+1}\right),
\end{equation}
the inequality \eqref{B.eq6} can be written as
\begin{equation}
\label{B.eq7}
  \left\langle\bLambda^{k+1} + \frac{1}{\mu}(\bSigma^{k} - \bSigma^{k+1}),\bY - \bY^{k+1}\right\rangle\leq 0,\quad \forall\; \bY\in \cP_{+} \textrm{ with } \Cond(\bY)\leq \kappa_n.
\end{equation}
Let $\bY = \bY^{k+1}$ in \eqref{B.eq5} and $\bY = \hat{\bY}^*$ in \eqref{B.eq6}, we can get
\begin{equation*}
  \left\langle\bY^{k+1}-\hat{\bY}^*,\hat{\bLambda}^*\right\rangle\leq 0 \textrm{ and }   \left\langle\bLambda^{k+1} + \frac{1}{\mu}(\bSigma^{k} - \bSigma^{k+1}),\hat{\bY}^* - \bY^{k+1}\right\rangle\leq 0,
\end{equation*}
from which the following inequality can be easily derived
\begin{equation}
\label{B.eq8}
\left\langle(\bLambda^{k+1} - \hat{\bLambda}^{*}) + \frac{1}{\mu}(\bSigma^{k} - \bSigma^{k+1}), \bY^{k+1} - \hat{\bY}^*\right\rangle\geq 0.
\end{equation}

The first order optimality condition for updating $\bSigma$ gives
\begin{eqnarray}
  0&\in& \left(\bSigma^{k+1} - \wbSigma_n\right)_{ij} + \bLambda^{k}_{ij} +  \frac{1}{\mu}\left(\bSigma^{k+1} - \bY^{k+1}\right)_{ij} + \lambda\partial|\bSigma^{k+1}_{ij}|, \quad i\neq j, \label{B.eq9} \\
  0&=& \left(\bSigma^{k+1} - \wbSigma_n\right)_{ii} + \bLambda^{k}_{ii} +  \frac{1}{\mu}\left(\bSigma^{k+1} - \bY^{k+1}\right)_{ii}, \label{B.eq10}
\end{eqnarray}
where $i = 1, 2, \cdots, p$ and $j = 1, 2, \cdots, p$. With \eqref{B.eqUpLmd}, \eqref{B.eq9} and  \eqref{B.eq10} can be written as
\begin{eqnarray}
   \frac{1}{\lambda}\left(\wbSigma_n -\bSigma^{k+1}  - \bLambda^{k+1}\right)_{ij} &\in& \partial|\bSigma^{k+1}_{ij}| , \quad i\neq j, \label{B.eq11} \\
   \left(\bSigma^{k+1} - \wbSigma_n\right)_{ii} + \bLambda^{k+1}_{ii} &=& 0, \label{B.eq12}
\end{eqnarray}
respectively. By the monotone property of subdifferential, combining \eqref{B.eq2}, \eqref{B.eq3}, \eqref{B.eq11} and \eqref{B.eq12} gives
\begin{equation}\label{B.eq16}
  \left\langle\bSigma^{k+1} - \hat{\bSigma}^{*}, (\hat{\bSigma}^{*} - \bSigma^{k+1}) + (\hat{\bLambda}^{*} - \bLambda^{k+1})\right\rangle\geq 0.
\end{equation}
With \eqref{B.eq8} and \eqref{B.eq16}, we can get
\begin{eqnarray}
  \left\|\bSigma^{k+1} - \hat{\bSigma}^{*}\right\|_F^2 & \leq & \left\langle\bSigma^{k+1} - \hat{\bSigma}^{*},\hat{\bLambda}^{*} - \bLambda^{k+1}\right\rangle + \left\langle\bLambda^{k+1} - \hat{\bLambda}^{*},\bY^{k+1} - \hat{\bY}^*\right\rangle\nonumber\\
   & & + \frac{1}{\mu}\left\langle\bSigma^{k} - \bSigma^{k+1},\bY^{k+1} - \hat{\bY}^*\right\rangle.\label{B.eq15}
\end{eqnarray}
Since $\hat{\bY}^* = \hat{\bSigma}^{*}$, with the updating formula of $\bLambda$ the inequality \eqref{B.eq15} can be equivalently written as
\begin{eqnarray}
  \left\|\bSigma^{k+1} - \hat{\bSigma}^{*}\right\|_F^2 - \left\langle\bSigma^{k} - \bSigma^{k+1},\bLambda^{k} - \bLambda^{k+1}\right\rangle
 &\leq& \frac{1}{\mu} \left\langle\bSigma^{k} - \bSigma^{k+1},\bSigma^{k+1} - \hat{\bSigma}^{*}\right\rangle \nonumber\\
   & &+\mu \left\langle\bLambda^{k} - \bLambda^{k+1},\bLambda^{k+1} - \hat{\bLambda}^{*}\right\rangle.\label{B.eq16}
\end{eqnarray}
Substituting the equalities $\bSigma^{k+1} - \hat{\bSigma}^{*} = \bSigma^{k+1} - \bSigma^{k} + \bSigma^{k} - \hat{\bSigma}^{*} $ and $\bLambda^{k+1} - \hat{\bLambda}^{*} = \bLambda^{k+1} - \bLambda^{k} +\bLambda^{k} - \hat{\bLambda}^{*}$ into the right-hand side of \eqref{B.eq16} leads to
\begin{eqnarray}
\frac{1}{\mu} \left\langle\bSigma^{k} - \bSigma^{k+1},\bSigma^{k} - \hat{\bSigma}^{*}\right\rangle &+& \mu \left\langle\bLambda^{k} - \bLambda^{k+1},\bLambda^{k} - \hat{\bLambda}^{*}\right\rangle\nonumber\\
&\geq& \left\|\bSigma^{k+1} - \hat{\bSigma}^{*}\right\|_F^2 - \left\langle\bSigma^{k} - \bSigma^{k+1},\bLambda^{k} - \bLambda^{k+1}\right\rangle\nonumber\\
 & & +  \frac{1}{\mu}\left\|\bSigma^{k+1} - \bSigma^{k}\right\|_F^2 + \mu \left\|\bLambda^{k+1} - \bLambda^{k}\right\|_F^2.\label{B.eq17}
\end{eqnarray}
Let $\bH^{k} =\begin{bmatrix}
                \bSigma^k \\
                \bLambda^k \\
              \end{bmatrix}$,
$\bH^{*} = \begin{bmatrix}
                \hat{\bSigma}^* \\
                \hat{\bLambda}^*\\
              \end{bmatrix}$, and
$\bW = \begin{bmatrix}
   \frac{1}{\mu}\bI_p &   \\
     & \mu \bI_p \\
 \end{bmatrix}
$, the inequality \eqref{B.eq17} can be transformed into
\begin{eqnarray}
\left\langle \bH^{k} - \bH^{k+1}, \bH^{k}- \bH^{*}\right\rangle_{\bW} &\geq& \left\|\bH^{k} - \bH^{k+1}\right\|_{\bW}^2 + \left\|\bSigma^{k+1} - \hat{\bSigma}^{*}\right\|_F^2\nonumber\\
& &- \left\langle\bSigma^{k} - \bSigma^{k+1},\bLambda^{k} - \bLambda^{k+1}\right\rangle,\label{B.eq18}
\end{eqnarray}
where $\left\|\cdot\right\|_{\bW}^2$ is the weighted norm, that is, $\left\|\bA\right\|_{\bW}^2 = \left\langle \bA,\bA\right\rangle_{\bW} = \left\langle \bA,\bW\bA\right\rangle$ for matrix $\bA$.
By the following equality
\begin{equation*}
  \left\|\bH^{k+1} - \bH^{*}\right\|_{\bW}^2 = \left\|\bH^{k} - \bH^{k+1}\right\|_{\bW}^2 - 2\left\langle \bH^{k} - \bH^{k+1}, \bH^k - \bH^*\right\rangle + \left\|\bH^{k} - \bH^{*}\right\|_{\bW}^2,
\end{equation*}
we have
\begin{eqnarray}
  \left\|\bH^{k} - \bH^{*}\right\|_{\bW}^2 -  \left\|\bH^{k+1} - \bH^{*}\right\|_{\bW}^2 &=& 2\left\langle \bH^{k} - \bH^{k+1}, \bH^k - \bH^*\right\rangle - \left\|\bH^{k} - \bH^{k+1}\right\|_{\bW}^2 \nonumber\\
   &\geq& \left\|\bH^{k} - \bH^{k+1}\right\|_{\bW}^2 + 2\left\|\bSigma^{k+1} - \hat{\bSigma}^{*}\right\|_F^2\nonumber\\
& &- 2\left\langle\bSigma^{k} - \bSigma^{k+1},\bLambda^{k} - \bLambda^{k+1}\right\rangle.\label{B.eq19}
\end{eqnarray}
If we substitute the index $k+1$ in \eqref{B.eq11} and \eqref{B.eq12} with $k$ and using the monotone property of subdifferential at $\bSigma^{k+1}$ and $\bSigma^{k}$ , we can get
\begin{equation*}
  \left\langle\bSigma^{k} - \bSigma^{k+1}, (\bSigma^{k+1} - \bSigma^{k}) + (\bLambda^{k+1} - \bLambda^{k})\right\rangle\geq 0,
\end{equation*}
which implies
\begin{equation}\label{B.eq20}
  -\left\langle\bSigma^{k} - \bSigma^{k+1}, \bLambda^{k} - \bLambda^{k+1}\right\rangle\geq  \left\|\bSigma^{k+1} - \bSigma^{k}\right\|_F^2 \geq 0.
\end{equation}
Thus, with \eqref{B.eq20} and \eqref{B.eq19} we can assert that
\begin{equation}
\label{B.eq21}
  \left\|\bH^{k} - \bH^{*}\right\|_{\bW}^2 -  \left\|\bH^{k+1} - \bH^{*}\right\|_{\bW}^2\geq \left\|\bH^{k} - \bH^{k+1}\right\|_{\bW}^2.
\end{equation}

According to \eqref{B.eq21}, we can easily get that the sequence $\left\{\left\|\bH^{k} - \bH^{*}\right\|_{\bW}^2\right\}$ is monotonically non-increasing and
\begin{equation}\label{B.eq22}
  \sum_{k=0}^{\infty}\left\|\bH^{k+1} - \bH^{k}\right\|_{\bW}^2 \leq \left\|\bH^{0} - \bH^{*}\right\|_{\bW}^2,
\end{equation}
which also implies $\lim_{k\rightarrow\infty}\left\|\bH^{k} - \bH^{k+1}\right\|_{\bW}^2 = 0$
and the sequence $\left\{\bH^{k}\right\}$ lies in a compact set.
By the denotation of $\bH^k$, we have $\bSigma^{k} - \bSigma^{k+1}\rightarrow 0 \textrm{ and }\bLambda^{k} - \bLambda^{k+1}\rightarrow 0$.
With \eqref{B.eq77}, we also have $\bSigma^{k} - \bY^{k} \rightarrow 0$. The compactness of $\left\{\bH^{k}\right\}$ ensures the existence of the limit point
$\bar{\bH} = \begin{bmatrix}
                \bar{\bSigma} \\
                \bar{\bLambda} \\
              \end{bmatrix}$ for a subsequence $\left\{\bH^{k_i}\right\}$. Since $\bSigma^{k} - \bY^{k} \rightarrow 0$, we may claim that $\left(\bar{\bSigma},\bar{\bY},\bar{\bLambda}\right)$ is a limit point of the sequence $\left\{\left(\bSigma^{k},\bY^{k},\bLambda^{k}\right)\right\}$ with $\bar{\bY}:= \bar{\bSigma}$ being a limit of the subsequence $\left\{\bY^{k_i}\right\}$. With the above discussion and the relationships \eqref{B.eq7}, \eqref{B.eq11} and \eqref{B.eq12},  the limit point $\left(\bar{\bSigma},\bar{\bY},\bar{\bLambda}\right)$ satisfies
\begin{eqnarray*}
   \frac{1}{\lambda}\left(\wbSigma_n -\bar{\bSigma}  - \bar{\bLambda}\right)_{ij} &\in& \partial|\bar{\bSigma}_{ij}| , \quad i\neq j, \label{B.eq11} \\
   \left(\bar{\bSigma} - \wbSigma_n\right)_{ii} + \bar{\bLambda}_{ii} &=& 0, \label{B.eq12}
\end{eqnarray*}
with $i = 1, 2, \cdots, p$ and $j = 1, 2, \cdots, p$, and
\begin{equation*}
\label{B.eq7}
  \left\langle\bar{\bLambda},\bY - \bar{\bY}\right\rangle\leq 0,\quad \forall\; \bY\in \cP_{+} \textrm{ with } \Cond(\bY)\leq \kappa_n.
\end{equation*}
This means given any starting point the limit $\left(\bar{\bSigma},\bar{\bY},\bar{\bLambda}\right)$ is an optimal solution to the optimization problem \eqref{eq.eqvamodel}, which completes the proof.
\end{proof}

\end{document}